\documentclass[lettersize,journal]{IEEEtran}
\usepackage{amsmath,amsfonts}
\usepackage{algorithmic}
\usepackage{algorithm}
\usepackage{array}
\usepackage[caption=false,font=normalsize,labelfont=sf,textfont=sf]{subfig}
\usepackage{textcomp}
\usepackage{stfloats}
\usepackage{url}
\usepackage{verbatim}
\usepackage{graphicx}
\usepackage{cite}

\usepackage{xcolor}
\usepackage{graphicx}
\usepackage{caption} 

\usepackage{multirow}
\usepackage{booktabs}

\begin{document}

\title{Towards Generalizability to Tone and Content Variations in the Transcription of Amplifier Rendered Electric Guitar Audio }

\author{Yu-Hua Chen$^{1}$, Yuan-Chiao Cheng$^{2}$, Yen-Tung Yeh$^{1}$, Jui-Te Wu$^{2}$, Jyh-Shing Roger Jang$^{1}$ and Yi-Hsuan Yang$^{1}$\\
\vspace{2mm}
        $^{1}$ National Taiwan University \;\;\; $^{2}$ Positive Grid \\

\thanks{}

}



\maketitle

\begin{abstract}

Transcribing electric guitar recordings is challenging due to the scarcity of diverse datasets and the complex tone-related variations introduced by amplifiers, cabinets, and effect pedals. To address these issues, we introduce EGDB-PG, a novel dataset designed to capture a wide range of tone-related characteristics across various amplifier-cabinet configurations. In addition, we propose the Tone-informed Transformer (TIT), a Transformer-based transcription model enhanced with a tone embedding mechanism that leverages learned representations to improve the model’s adaptability to tone-related nuances. Experiments demonstrate that TIT, trained on EGDB-PG, outperforms existing baselines across diverse amplifier types, with transcription accuracy improvements driven by the dataset’s diversity and the tone embedding technique. Through detailed benchmarking and ablation studies, we evaluate the impact of tone augmentation, content augmentation, audio normalization, and tone embedding on transcription performance. This work advances electric guitar transcription by overcoming limitations in dataset diversity and tone modeling, providing a robust foundation for future research.
\end{abstract}


\section{Introduction}
\label{introduction}

Automatic music transcription (AMT) has attracted considerable interest, achieving significant progress through deep learning techniques \cite{emiya2009multipitch,hawthorne2021sequence, yan2021skipping, ou2022exploring,hawthorne2017onsets,kong2021high, hawthorne2018enabling}. A robust AMT system is essential for various Music Information Retrieval (MIR) tasks, including beat tracking, chord recognition, and performance analysis. Furthermore, accurate transcription models facilitate automatic MIDI generation, e.g., enabling a  cycle of audio-to-score and score-to-audio conversion in the case of piano music \cite{hawthorne2018enabling}.

Thanks to extensive datasets like MAPS (65 hours) \cite{emiya2009multipitch} and MAESTRO (over 200 hours) \cite{hawthorne2018enabling}, AMT has advanced significantly, especially in piano transcription.
These datasets are critical because they offer the diverse and voluminous data required to train neural network (NN) models, which thrive on large amounts of data. The richness of these datasets has been a key factor in achieving high transcription accuracy. The models used in this field are sequence-modeling architectures, such as recurrent neural networks (RNNs) \cite{hawthorne2017onsets} and Transformers \cite{toyama2023automatic, vaswani2017attention}. These architectures treat spectrogram-based inputs as sequential data similar to language tokens. This approach allows them to capture temporal relationships, identify the onset and frame (duration of note) of each note for pitch detection, and model note dependencies within a score, resulting in precise note sequence predictions.

\begin{figure}[t]
    \centering
    \subfloat[Clean]{%
        \includegraphics[width=0.45\linewidth]{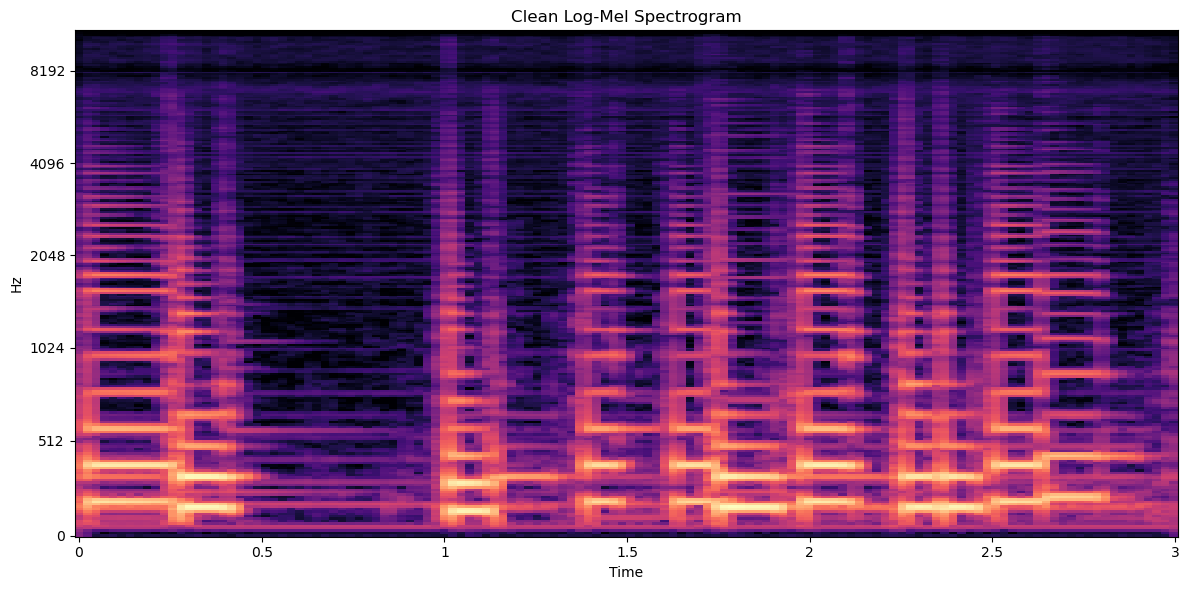}%
        \label{fig:out_domain_sub2}
    }
    \hfill
    \subfloat[High-gain]{%
        \includegraphics[width=0.45\linewidth]{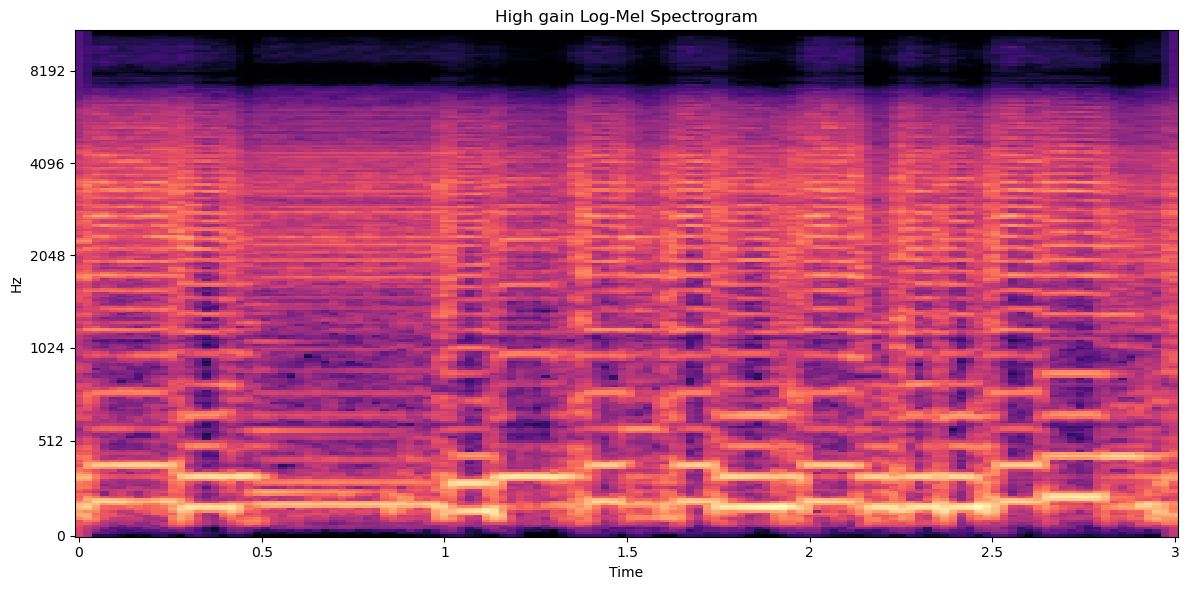}%
        \label{fig:out_domain_sub3}
    }

    \caption{Log-mel spectrograms of recordings of (a) clean audio and (b) audio rendered by a high-gain amplifier for the same musical content, demonstrating the significant alteration in frequency distribution and intensity. The high-gain processing introduces additional harmonic content and noise, illustrating the challenges faced by transcription models in generalizing across varied audio signals.}
    \label{fig:clean-high-gain}
\end{figure}

Electric guitars play a pivotal role in pop and modern music due to their versatile sound design and expressive potential, highlighting the importance of accurate transcription for educational initiatives, especially for beginners~\cite{wang2021soloist}. 
However, transcribing electric guitar audio presents distinct challenges, including limited data quantity, lack of tone diversity, complexities of tablature format, and expressive playing techniques, which are further detailed in Section~\ref{AGT}.
Unlike acoustic instruments like the piano, an electric guitar generates sound via a direct input (DI) signal, a raw electrical output from its pickups that captures string vibrations, which is then transformed by analog or digital devices such as amplifiers, effect pedals, and plugins. These devices shape the DI signal through combinations of amplifier heads, cabinets, effect pedals, and digital plugins, creating virtually unlimited ``presets'' that define the overall signal chain. In guitarist parlance, the resulting tone or voice refers to the distinctive sonic character of the guitar, enabling everything from clean and bright timbres to aggressive and distorted textures, such as warm jazz tones, gritty rock edges, or searing lead sounds. We define the processed output as ``wet audio'', the DI signal after being rendered with a given preset, introducing gain stages, harmonic distortion, and other effects that result in a rich and varied sonic output. Consequently, the transcription of electric guitar audio is complicated by this inherent tone diversity, characterized by continuous spectral and temporal variations and overlapping harmonics, as illustrated in Figure~\ref{fig:clean-high-gain}. These tone-related nuances pose substantial generalization challenges for transcription models, a difficulty further exacerbated by the electric guitarist’s pursuit of an ideal tone, which is a defining endeavor driven by the instrument’s versatility and the subjective artistry of sculpting a signature sound.

Our objective is to advance electric guitar transcription by focusing on \textit{amplifier-rendered audio}, a task that requires mapping diverse audio variations, processed through various effects, to a unified score, framing it as a many-to-one problem. We address this challenge through a series of contributions that explore the potential of expanding the duration and heterogeneity of training data with diverse playing styles and genres to improve multi-pitch prediction accuracy in automatic electric guitar transcription systems, mirroring advancements in piano transcription. We also investigate whether leveraging a large-scale dataset of amplifier-rendered electric guitar tones with varied signal processing and distortion profiles can enhance transcription performance, and how a transcription model can be designed or trained to accurately handle diverse tone-related conditions, ranging from low-gain to high-gain amplifier sounds.

To evaluate the impact of expanded and diverse training data on multi-pitch prediction accuracy, we introduce \textbf{EGDB-PG}, a dataset created by rendering the EGDB dataset \cite{chen2022towards} with BiasFX2 plugins from Positive Grid, a world-leading guitar amplifier and plugin company. This dataset addresses the scarcity of publicly available guitar datasets that feature effect-rendered audio with alignment labels, a critical limitation for training robust transcription models. By incorporating 256 unique combinations of 16 amplifier heads and 16 cabinets, EGDB-PG captures a broad spectrum of tone-related variations, from low-gain to high-gain amplifier tones, providing a comprehensive resource to test and improve model generalizability. We make EGDB-PG public to support future research  at our project page.\footnote{\url{https://ss12f32v.github.io/Guitar-Transcription-with-Amplifier/}}

As the dataset is expanded in a tone-varied manner, we explore whether leveraging large-scale tone datasets enhances transcription performance and how models can be designed to handle diverse tone conditions. To this end, we integrate a tone embedding mechanism into our transcription model, drawing inspiration from query-based music source separation techniques~\cite{lee2019audio, liu2024separate, chen2022zero}. Following~\cite{chen2024towards}, we extract tone embeddings from amplifier-rendered audio to condition the model, improving its adaptability to tone differences. We modify the hierarchical Frequency-Time Transformer (hFT-Transformer)~\cite{toyama2023automatic}, which consists of two-dimensional Transformer modules operating along frequency and time axes, by incorporating a cross-attention mechanism with tone embeddings. We name this enhanced model as the Tone-informed Transformer (TIT) in this paper.

We evaluate this approach through a series of experiments using the EGDB-PG dataset. Initially, we conduct ablation studies on the test set of EGDB-PG, assessing the impact of various training strategies, including content augmentation, tone augmentation, tone embedding, and audio normalization. These studies identify the optimal settings that maximize transcription accuracy for electric guitar audio. Using the best-performing settings, we then compare our tone-embedding-enhanced model against different baseline models, demonstrating improved performance across various amplifier types and confirming the value of large-scale tone datasets and tone-aware modeling. 
Finally, to validate the robustness of our approach, we evaluate the model on an unseen, out-of-domain amplifier tone dataset that includes tones from Neural DSP, a different guitar plugin company. These tones are not present in our training data, which further supports the findings from our ablation studies and highlights the practical importance of effect modeling techniques in improving transcription accuracy across diverse tone-related conditions.

The rest of this paper is organized as follows. Section~\ref{background} describes the challenges of electric guitar transcription and defines the scope discussed in this paper. Section~\ref{proposed_dataset} presents the proposed EGDB-PG dataset. Section~\ref{methodology} describes the tone representation used in our work. Section~\ref{TIT_arch} details the architecture of our proposed Tone-informed Transformer (TIT) model and the training techniques employed. Section~\ref{sec:expereiment} presents the experimental results, and Section~\ref{conclusion} concludes the work. Examples of transcription result, including those for the out-domain test data, and additionally in-the-wild guitar recordings from YouTube, can be found at our project page.


\section{Background}
\label{background}
\subsection{Automatic Guitar Transcription}
\label{AGT}
 AMT for electric guitars presents unique challenges compared to piano transcription, a well-established domain in MIR. While piano transcription has benefited from significant advancements in neural networks and large-scale datasets, guitar transcription relatively underexplored. We identify four primary challenges in automatic electric guitar transcription: 1) limited data quantity in datasets, 2) lack of tone diversity, 3) complexities of tablature format stemming from guitar fretboard design, and 4) expressive playing techniques. These challenges are detailed below, followed by a discussion of our efforts to address the first two, while the latter two require further investigation that is beyond the scope of this work.

\subsubsection{Limited Data Quantity in Datasets}
The quantity of data available for training guitar transcription models is notably limited compared to piano transcription, which leverages extensive datasets such as MAPS \cite{emiya2009multipitch} and MAESTRO \cite{hawthorne2018enabling}. These piano datasets provide the scale and variety necessary to train data-intensive neural networks effectively, as seen in models like the Onsets and Frames (OAF) series~\cite{hawthorne2017onsets,kong2021high,hawthorne2018enabling}. In contrast, guitar datasets suffer from a significant disparity in duration. For example, GuitarSet~\cite{xi2018guitarset} offers only 360 clips of solo acoustic guitar recordings, totaling a few hours, while EGDB~\cite{chen2022towards} includes 240 DI samples for electric guitars. Synthesized datasets, such as Synthesized Lakh (Slakh)\cite{manilow2019cutting} with over 5,000 MIDI guitar tracks, and SynthTab\cite{zang2024synthtab} derived from DadaGP scores~\cite{sarmento2021dadagp}, provide larger volumes but lack the realism of human performances. This scarcity of large-scale, real-performance data hinders the development of robust transcription models for guitars.

\subsubsection{Lack of Tone Diversity}
A critical limitation in existing guitar datasets is the lack of tone diversity, particularly for electric guitars where tone shaped by amplifiers, effects chains, and recording setups is a defining characteristic. While piano datasets benefit from the consistent acoustic properties of the instrument, electric guitar audio exhibits a vast range of tone variations that complicate transcription. For instance, EGDB~\cite{chen2022towards} renders its DI samples with five distinct amplifier configurations, but this represents only a narrow subset of possible tones. The absence of datasets capturing a broader spectrum of amplifier-rendered and effect-laden audio impedes the development of models capable of generalizing across diverse tone conditions.

\subsubsection{Complexities of Tablature Format from Guitar Fretboard Design}
Unlike piano music, which maps directly to discrete pitches and is often represented as piano rolls (a 2D format of keys $\times$ frames), guitar music is traditionally notated in the tablature format, reflecting the instrument’s fretboard design. This notation specifies strings and frets rather than absolute pitches, introducing complexity for transcription models. A single pitch can often be played on multiple strings, creating ambiguity in mapping notes to positions. For example, TabCNN~\cite{wiggins2019guitar} models tablature as a 2D image (strings $\times$ frets, frames), but this approach struggles with disambiguating identical pitches across strings, adding a layer of difficulty beyond piano transcription.

\subsubsection{Expressive Playing Techniques}
The guitar’s fretboard enables expressive techniques such as bending and sliding, which produce continuous pitch variations rather than the discrete notes typical of pianos. These techniques result in dynamic pitch contours that challenge conventional transcription models designed for fixed pitches. While piano transcription leverages sequence-modeling architectures like RNNs\cite{hawthorne2017onsets} or Transformers\cite{toyama2023automatic,vaswani2017attention} to capture temporal relationships and predict note sequences, guitar audio demands models capable of handling non-discrete pitch transitions---a capability not yet fully developed in existing frameworks \cite{su2019tent, huang2023note}.

\subsubsection{Current Efforts and Remaining Gaps}

Recent efforts have begun to address the challenges of limited data quantity and tone diversity, though significant gaps remain. For data quantity, Riley et al.~\cite{riley2024high} explored pre-training transcription models on the MAESTRO dataset before fine-tuning them for acoutic guitar audio transcription, but this approach has not specifically tackled the challenges of electric guitar audio. To enhance tone diversity, studies such as Abreu et al.~\cite{abreu2024leveraging} have expanded GuitarSet and synthesized GuitarPro data using randomly selected effect configurations~\cite{pedroza2022egfxset}, while our previous work~\cite{chen2022towards} trained models on five different types of amplifier-rendered audio. Despite these efforts, transcription performance on large-scale, effect-rendered audio relatively underexplored. Moreover, the challenges of tablature format and expressive playing techniques require further investigation. While some studies have adapted piano-roll representations to prioritize pitch clarity over tablature specificity (e.g., Riley et al.~\cite{riley2024high} employed polyphonic score alignment to refine datasets), these approaches do not fully capture the guitar’s fretboard-specific notation intricacies or its expressive techniques. Addressing these challenges demands novel modeling strategies and richer datasets tailored to the guitar’s unique characteristics.



\begin{table*}[h]
    \centering
    \begin{tabular}{l l l r r r}
        \hline
        Dataset & Instrument & Live Recording & \multicolumn{1}{r}{Duration (Clean)} & \multicolumn{1}{r}{\# of Presets} & \multicolumn{1}{r}{Duration (Total)} \\
        \hline
        EGDB-PG & Electric guitar & True & 2 hrs & 256 & 514 hrs \\
        \hline
        GuitarSet \cite{xi2018guitarset} & Acoustic guitar & True & 3 hrs & 1 & 3 hrs \\
        EGDB \cite{chen2022towards} & Electric guitar & True & 2 hrs & 5 & 12 hrs \\
        IDMT-SMT-Guitar \cite{stein2010automatic} & Acoustic \& Electric guitar & True & 35 mins & 11 & 19 hrs \\
        SynthTab \cite{zang2024synthtab} & Acoustic \& Electric guitar & False & n/a & 23 & 13113 hrs \\
        \hline
        Maestro \cite{hawthorne2018enabling} & Piano & True & 199 hrs & 1 & 199 hrs \\
        \hline
    \end{tabular}
    \caption{Comparison of datasets for electric guitar transcription, detailing instrument type, recording methods (True for live recordings, False for synthesized audio), and duration metrics for clean and effect-rendered audio. The '\# of Presets' indicates the number of unique effect chains, where each chain includes one preset, such as an amplifier (head and cabinet combination) counted as a single preset. The piano dataset Maestro is included to highlight the significant disparity in clean duration between piano and guitar datasets, underscoring differing data needs for transcription tasks.}
    \label{tab:dataset_comparison}
\end{table*}

\subsection{Tone informed MIR}
\label{Q_B_MIR}
Recent advancements in query-based music source separation\cite{lee2019audio, liu2024separate, chen2022zero}, a well-established task in Music Information Retrieval (MIR), have broadened the scope of condition-based audio processing. This approach employs a query—represented as an audio clip, text, or visual input—to condition a model, typically encoding the query as a latent vector within a continuous latent space. By isolating specific audio elements based on this condition, query-based source separation enables zero-shot capabilities \cite{lee2019audio, liu2024separate, chen2022zero}, provided the latent vector effectively encapsulates the query’s defining features. Such techniques offer a flexible framework for manipulating complex audio signals, inspiring applications beyond traditional separation tasks.

In the realm of neural guitar amplifier modeling, our previous work \cite{chen2024towards} pioneered a method that conditions a temporal convolutional network with a tone embedding to represent amplifier-specific tone characteristics. This embedding is derived from a tone encoder trained via contrastive learning, enabling the model to generalize to unseen amplifier tones. By extracting embeddings from novel guitar clips using the pre-trained encoder, this approach captures nuanced spectral and temporal properties, facilitating zero-shot tone modeling. These developments in conditioned audio processing directly inform our transcription model, which adapts tone embeddings to address the tone diversity of electric guitar audio.

\begin{figure*}[ht]
    \centering
    \subfloat[Pianoroll]{%
        \includegraphics[width=0.19\linewidth]{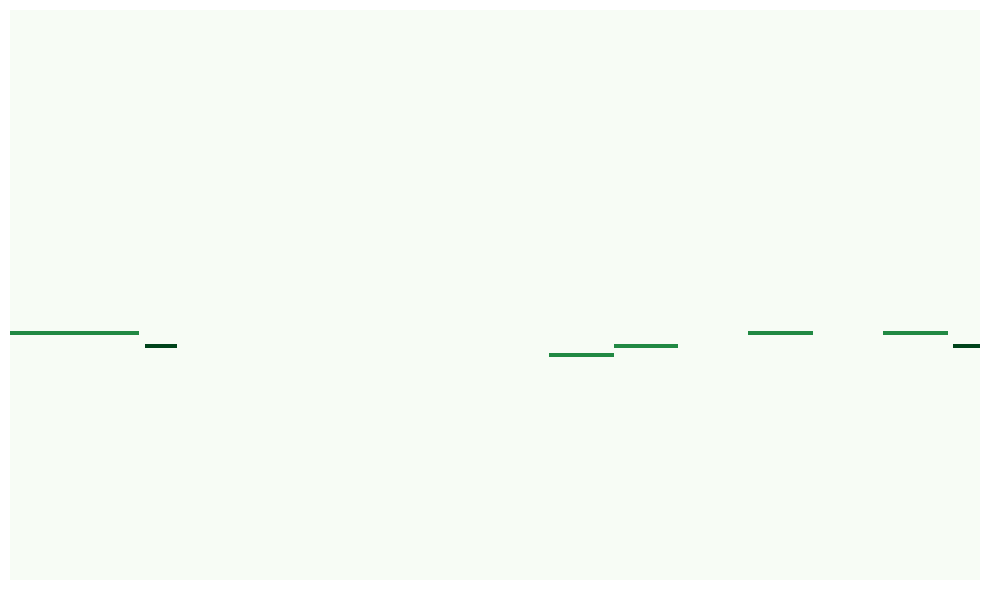}
        \label{fig:sub1}
    }
    \hfill
    \subfloat[Clean]{%
        \includegraphics[width=0.19\linewidth]{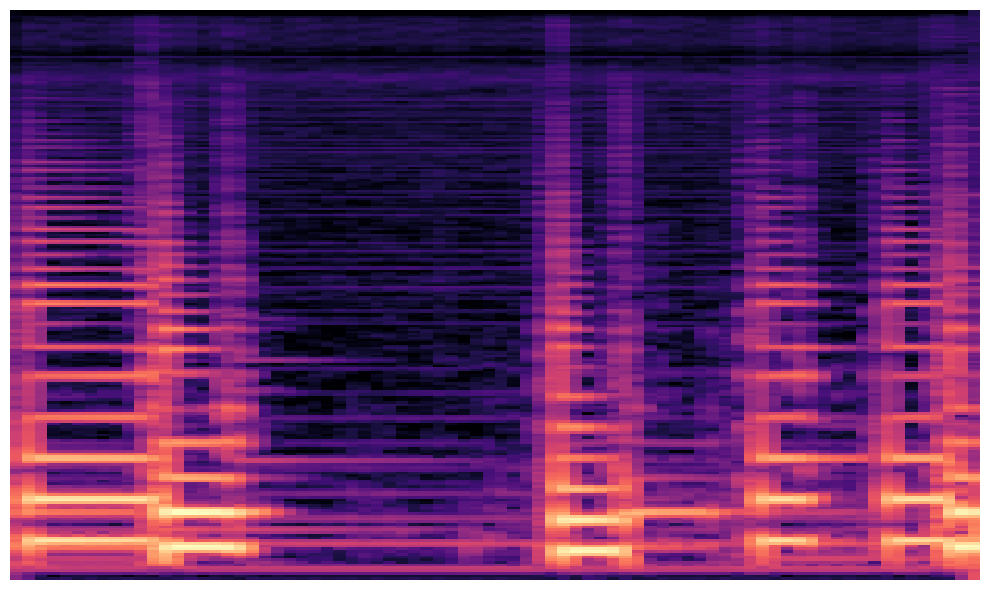}
        \label{fig:sub2}
    }
    \hfill
    \subfloat[Low-gain]{%
        \includegraphics[width=0.19\linewidth]{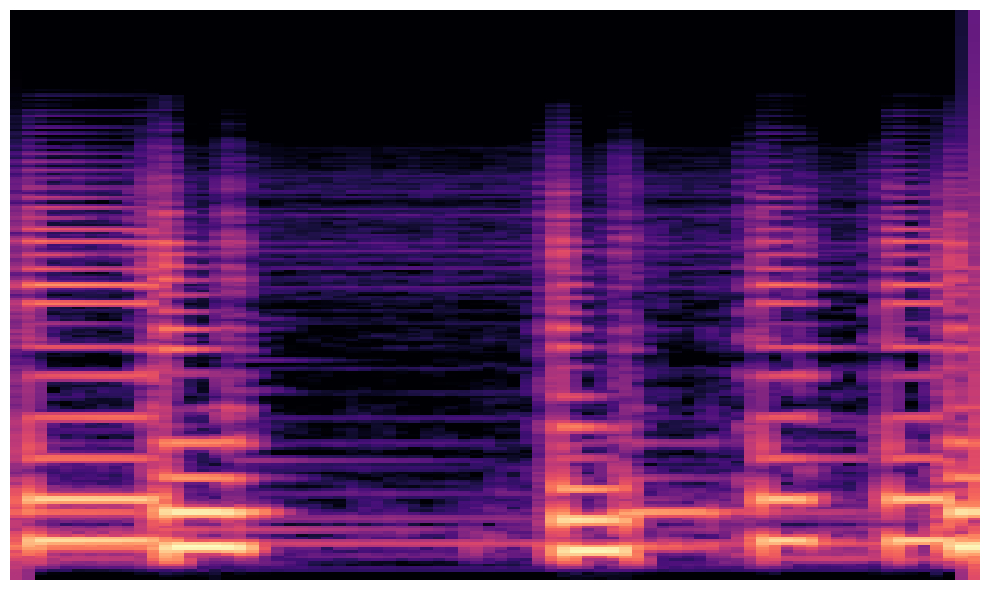}
        \label{fig:sub3}
    }
    \hfill
    \subfloat[Crunch]{%
        \includegraphics[width=0.19\linewidth]{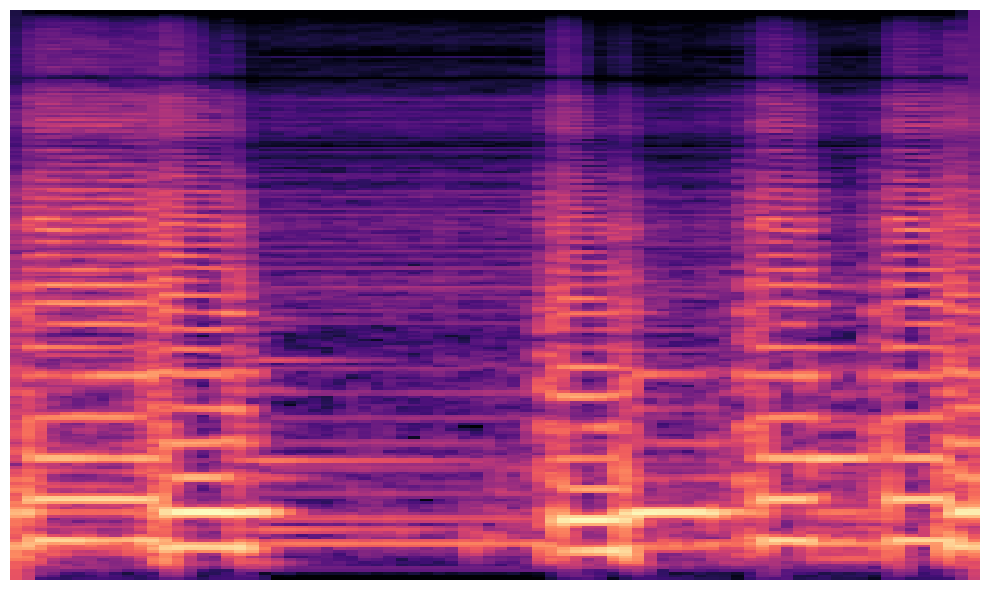}
        \label{fig:sub4}
    }
    \hfill
    \subfloat[High-gain]{%
        \includegraphics[width=0.19\linewidth]{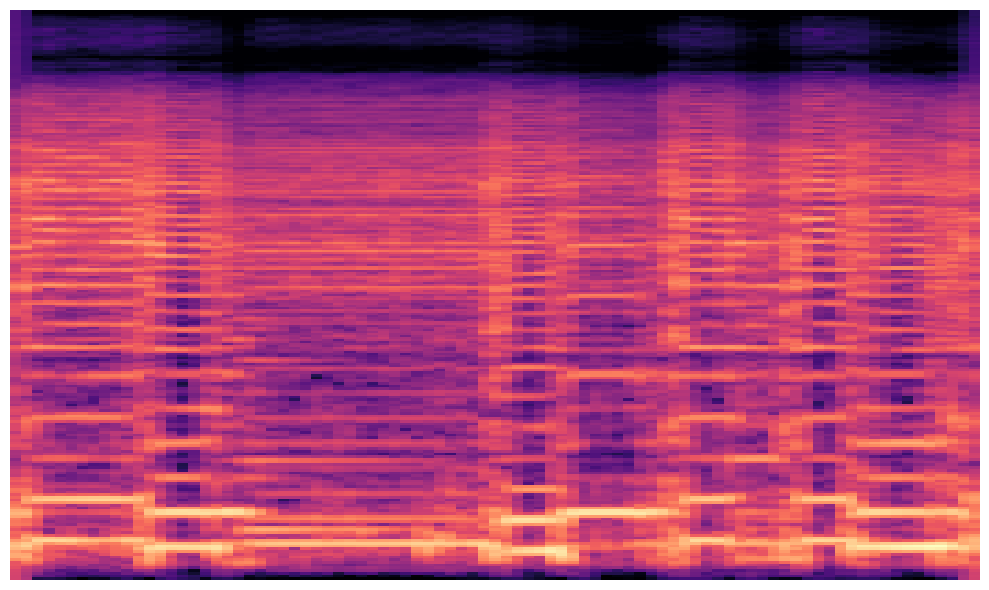}
        \label{fig:sub5}
    }

    \caption{Visualization of audio representations with varying amplifier presets from the EGDB-PG dataset. (a) Piano roll displaying the note labels of the audio content. (b–e) Log-mel spectrograms of the same audio content processed with one setting from each amplifier type category defined in EGDB-PG: (b) clean (DI) audio, (c) low-gain, (d) crunch, and (e) high-gain amplifiers. These visualizations demonstrate how amplifier settings modify the spectral characteristics of the audio, including differences in harmonic and non-harmonic growth as gain increases, while preserving the underlying musical content.}
    \label{fig:dataset_demonstration}
\end{figure*}

\section{EGDB-PG}
\label{proposed_dataset}

As highlighted in Section \ref{AGT}, existing guitar datasets suffer from insufficient tone variation, which has hindered the development and validation of robust electric guitar transcription models. To address this gap, we begin with the EGDB dataset, a widely recognized resource in guitar audio research, which focuses on electric guitar audio with limited effect processing, incorporating renderings from five amplifiers \cite{chen2022towards}. We augmented all audio clips from the EGDB dataset by re-rendering the clean audio with the Positive Grid BiasFX2 plugin, generating 256 unique presets through combinations of 16 amplifier heads and 16 cabinets. This process produced EGDB-PG, an expanded dataset offering a diverse tone palette to challenge transcription models and enhance their generalizability. Each preset is classified into one of three amplifier types, low-gain, crunch, and high-gain, based on the gain characteristics of the amplifier head, resulting in 96, 64, and 96 presets, respectively. These categories are visually represented in Figure \ref{fig:dataset_demonstration}, which illustrates log-mel spectrograms of a clean signal, low-gain, crunch, and high-gain amplifier renderings, demonstrating the tone diversity introduced by our augmentation.

For EGDB-PG, we partitioned the clean signals into training, validation, and testing sets with a 90/5/5 ratio. 
The rendered dataset, EGDB-PG, comprises 514 hours of audio, as detailed in Table \ref{tab:dataset_comparison}, providing an extensive resource to address prior dataset shortcomings. While EGDB-PG significantly enhances tone coverage, unannotated playing techniques like palm muting and ghost note may limit real-world applicability, suggesting future annotation efforts. This dataset enables our tone embedding model to generalize across 256 tone variations, as validated in Section V, directly supporting our research questions on tone diversity and model adaptability.

\section{Tone representation}
\label{methodology}
\subsection{Amplifier-Rendered Transcription Approaches}
\label{methodology:amplifier_render_transcription}

In this study, we introduce the \textbf{tone-informed transcription approach} to tackle the challenges of transcribing amplifier-rendered guitar audio, a task complicated by the diverse tone variations introduced by amplifiers and effects, as noted in Section~\ref{AGT}. This method conditions the transcription model on a tone representation \( \mathbf{c} \), which encapsulates the tone characteristics of the amplifier-rendered audio \( \mathbf{x}_r \).

Let \( \mathbf{x} \) denote the clean audio signal, and \( \mathbf{s} \) represent the corresponding musical score (e.g., a piano roll with onset, frame, and offset components) for this clean audio. The amplifier-rendered audio \( \mathbf{x}_r \) is obtained through a non-linear transformation \( g(\cdot) \), such that:
\begin{equation}
\mathbf{x}_r = g(\mathbf{x}),
\end{equation}
where \( g(\cdot) \) models the series of non-linear transformations introduced by the amplifier, including gain stages, harmonic distortion, and other effects. To account for the diversity of amplifier settings, we define a set of presets \( \Theta \), which represents a wide range of possible configurations (e.g., combinations of amplifier heads, cabinets, and effect parameters). In EGDB-PG, \( \Theta \) specifically comprises 256 presets, each representing a unique combination of amplifier settings. Each preset \( \theta \in \Theta \) corresponds to a specific configuration, and the rendered audio for a given preset is expressed as:
\begin{equation}
\mathbf{x}_{r,\theta} = g(\mathbf{x}, \theta).
\end{equation}
Thus, different presets \( \theta_1, \theta_2 \) produce distinct rendered audio signals \( \mathbf{x}_{r,\theta_1} \) and \( \mathbf{x}_{r,\theta_2} \), both of which map to the same underlying score \( \mathbf{s} \), framing the task as a many-to-one mapping problem.

The tone representation \( \mathbf{c} \) is derived from the amplifier-rendered audio \( \mathbf{x}_{r,\theta} \) using a pre-trained model \( f_{\text{tone}}(\cdot) \), which compresses the audio into a fixed-dimensional embedding:
\begin{equation}
\mathbf{c} = f_{\text{tone}}(\mathbf{x}_{r,\theta}).
\end{equation}
This tone embedding \( \mathbf{c} \), capturing the tone characteristics of \( \mathbf{x}_{r,\theta} \), is then provided as a conditioning input to the transcription model \( h_{\text{trans}}(\cdot) \). The model predicts the score components (onset, frame, and offset) from the rendered audio, conditioned on the tone embedding:
\begin{equation}
\hat{\mathbf{s}} = h_{\text{trans}}(\mathbf{x}_{r,\theta}, \mathbf{c}),
\end{equation}
where \( \hat{\mathbf{s}} \) denotes the predicted score, aligning with the dimensions of \( \mathbf{s} \). By leveraging tone embeddings, this approach enables the transcription model to adapt to the diverse tone properties of amplifier-rendered audio inputs, effectively addressing the challenges posed by varying presets \( \theta \in \Theta \).

\subsection{Tone Representation}
 In our approach, the tone encoder model is designed to compress and encode the tone or timbre information from a short audio clip (3–5 seconds in duration) into a single embedding vector. This tone embedding is intended to provide information about the current tone or timbre present in the audio and is used as an input to the transcription model.

The intuition behind this approach is that by representing tone variations as a single embedding, the transcription model might be able to map audio clips with different tone characteristics, originating from the same clean audio but rendered through diverse amplifier settings, to a identical piano-roll representation. This setting assumes that the encoded tone embedding can capture relevant tone information, helping the transcription process focus on the underlying musical content rather than tone variations.

\subsubsection{Contrastive learning}
Our objective is to train an encoder \( E \) that extracts tone- or style-related features from wet guitar audio signals while minimizing sensitivity to content-related information. This task involves achieving style-content disentanglement in the representation space. To this end, we adopt the self-supervised contrastive learning framework proposed in SimCLR \cite{chen2020simple}, originally developed for image representation learning \cite{caron2020unsupervised, assran2022maskedsiamesenetworkslabelefficient}, to train the audio encoder.

Positive pairs \( (\mathbf{x}_{r,\theta_1}, \mathbf{x'}_{r,\theta_1})_+ \) are defined as audio clips that share the same tone, determined by the preset \( \theta_1 \), but differ in their playing content, such as distinct musical phrases or notes. Negative pairs, denoted as \( (\mathbf{x}_{r,\theta_1}, \mathbf{x'}_{r,\theta_2})_- \), consist of audio clips with different tones, represented by distinct presets \( \theta_1 \) and \( \theta_2 \), regardless of whether the playing content is the same or different. Each audio clip is processed through an encoder \( f_{\text{tone}}(\cdot) \), which maps the input audio to an embedding representation \( \mathbf{c} = f_{\text{tone}}(\mathbf{x}_{r,\theta}). \), capturing the tone characteristics in a fixed-dimensional space. The training objective for the encoder is to maximize the similarity between embeddings of positive pairs, ensuring that clips with identical tones are closely aligned in the embedding space, while minimizing the similarity between embeddings of negative pairs, encouraging the model to distinguish between different tones effectively.


Training \( f_{\text{tone}}(\cdot) \) effectively requires a diverse dataset of wet guitar signals featuring a variety of tones and musical content. However, obtaining such a dataset is non-trivial. Guitar signals collected in the wild often lack annotations necessary to distinguish between tone- and content-related features. Alternatively, generating synthetic data via software tools like Pedalboard \cite{pedalboard2021} is limited by the diversity of tones available in open-source libraries.

To address this limitation, we leverage a collaboration with Positive Grid. We utilize a large collection of clean guitar signals as input to Positive Grid’s BiasFX2 to generate wet signals. By varying the combinations of amplifiers and effects pedals, we produce a dataset with extensive tone diversity. This approach enables the encoder to generalize across a wide spectrum of tone styles, ensuring robustness to variations in musical content.







The positive and negative pairs are designed to train the encoder \( f_{\text{tone}}(\cdot) \) to learn a tone-preserving projection that effectively captures tone characteristics in the embedding space. As training progresses, the cosine similarity between embeddings of clips with the same tone, defined as \( \text{sim}(\mathbf{c}, \mathbf{c}') = \frac{\mathbf{c}^\top \mathbf{c}'}{\|\mathbf{c}\| \|\mathbf{c}'\|} \), approaches 1:

\[
\text{sim}(f_{\text{tone}}(\mathbf{x}_{r,\theta_1}), f_{\text{tone}}(\mathbf{x'}_{r,\theta_1})) \rightarrow 1.
\]

Conversely, the similarity between embeddings of clips with different tones approaches 0:

\[
\text{sim}(f_{\text{tone}}(\mathbf{x}_{r,\theta_1}), f_{\text{tone}}(\mathbf{x'}_{r,\theta_2})) \rightarrow 0.
\]

This approach enforces the encoder to project clips with the same tone to similar points in the embedding space, while ensuring dissimilar pairs are mapped to distant regions. Consequently, the encoder learns tone-invariant features, achieving effective disentanglement of tone and content.

\begin{figure*}[htbp]
    \centering
     \includegraphics[width=\textwidth]{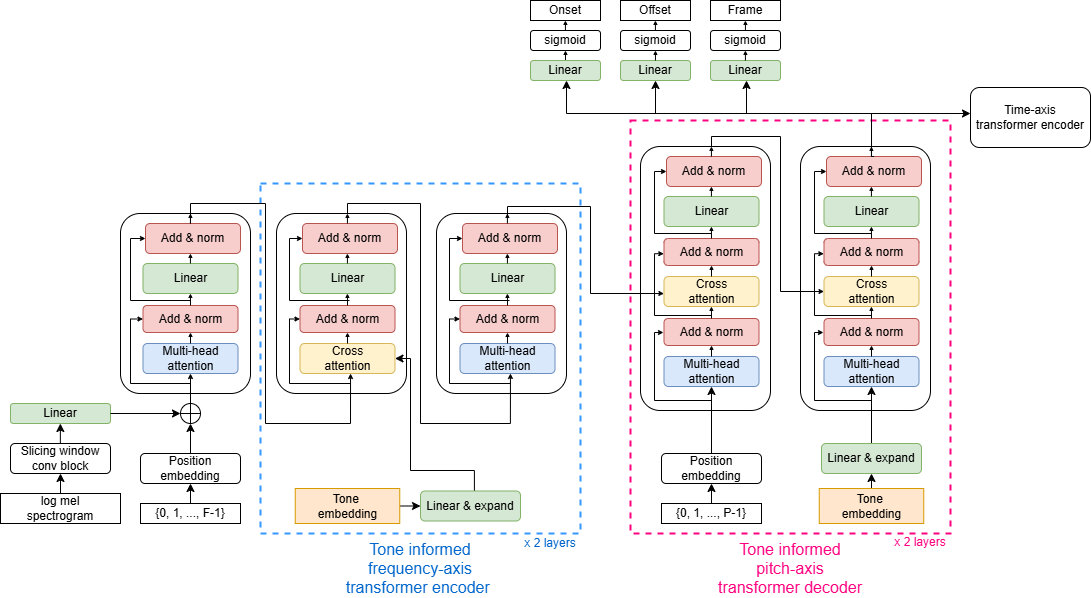}
    \caption{Architecture of the Tone-informed Transformer (TIT). The model features a frequency encoder with two modules: the tone-informed frequency-axis Transformer encoder, which processes spectrogram inputs along the frequency axis with tone embeddings integrated via cross-attention, and the tone-informed pitch-axis Transformer decoder, which refines pitch-related features. The frequency encoder generates predictions for note onsets, offsets, and frames, which are further processed by a time-axis Transformer encoder.}
    \label{fig:model_arch}
\end{figure*}

\section{Transcription System}
\label{TIT_arch}
This section introduces our TIT model, designed to enhance transcription accuracy for amplifier-rendered guitar performances. We begin by summarizing the hFT-Transformer~\cite{toyama2023automatic}, which forms the foundational architecture for TIT. Following the description of the hFT-Transformer, we detail the TIT architecture and the training techniques employed in our transcription system, including content augmentation and audio normalization.

\subsection{hFT-Transformer}

The hFT-Transformer is an encoder-decoder model tailored for hierarchical audio processing, operates in two stages to capture both spectral and temporal dependencies within the input log-mel spectrogram, making it well-suited for transcribing complex audio signals.

In the first stage, the model processes the log-mel spectrogram along the frequency axis. A convolutional block reduces feature dimensionality, followed by a Transformer encoder that models spectral relationships. Positional embeddings are applied along the frequency axis to preserve order, and a Transformer decoder maps the frequency dimension to the pitch dimension. The outputs of this stage, referred to as first output, include predictions for frame, onset and offset, aligning with the piano-roll representation.

In the second stage, the first output is further processed along the time axis. A Transformer encoder, equipped with time-based positional embeddings, captures temporal dependencies, yielding the final predictions as the second output. This hierarchical design balances spectral and temporal information, critical for accurate transcription of musical signals.



\subsection {Architecture of Tone-informed Transformer}

Amplifier effects typically introduce non-linear characteristics, such as harmonics and distortion, which predominantly manifest along the frequency axis in the frequency domain. Motivated by this observation, we propose a new model, named the Tone-informed Transformer (TIT), by modifying the hFT-Transformer architecture~\cite{toyama2023automatic}. Specifically, we inject tone information exclusively into the \textit{first stage} of the hFT-Transformer, where self-attention operates along the frequency axis. This design choice ensures that the model effectively captures the tone modifications induced by amplifiers, aligning with their spectral properties, while preserving the second-stage time-axis Transformer as in the original hFT-Transformer architecture.

The modified first stage, a tone-informed frequency-axis Transformer encoder, is illustrated in Figure~\ref{fig:model_arch}. The input log-mel spectrogram is first transformed into a feature representation and processed through a self-attention mechanism to capture frequency-domain dependencies. Subsequently, a cross-attention mechanism is applied, where the tone embedding serves as the query, and the encoded spectrogram acts as both the key and value. This cross-attention step is followed by another self-attention layer, forming a two-layer architecture for the frequency-axis encoder. The resulting frequency-axis feature map is then transformed from the number of mel bins to the number of pitches, aligning the output with the piano-roll-shaped representation required for transcription.

In the subsequent module of the first stage, a tone-informed pitch-axis Transformer decoder refines the pitch predictions by incorporating pitch-axis positional embeddings and tone embeddings. These embeddings are alternately used as keys in the cross-attention mechanism, enhancing the content derived from the frequency-axis encoder by integrating both pitch-specific and tone-specific characteristics. This process improves transcription accuracy by ensuring that the model accounts for tone variations while refining pitch information. The decoder’s output is passed through a linear transformation followed by a sigmoid activation function to produce logits for each pixel in the piano-roll representation, corresponding to frame, onset, and offset predictions.

For the second stage of TIT, the time-axis Transformer encoder, we retain the original hFT-Transformer architecture without injecting tone embeddings. This decision is based on the assumption that the amplifier tones considered in this work primarily affect the frequency axis of the spectrogram, with minimal impact along the time axis. 
This approach leverages cross-attention to adaptively integrate tone and pitch-related features across the frequency and pitch axes, enabling the model to better distinguish tone variations and improve transcription accuracy for amplifier-rendered audio.

We configure the model with the following parameters: 256 mel frequency bins, \( T = 100 \) time frames, a feature dimension of \( F = 256 \), and \( P = 49 \) pitches (ranging from pitch 40 to 88). The CNN uses \( C = 4 \) channels and a kernel size of \( K = 5 \). For the Transformer, the feed-forward network dimension is set to 256 with 4 attention heads. Each output head for onsets, offsets, and frames has a shape of \( P \times T \), corresponding to the number of pitches and time frames.

\subsection{Content Augmentation}
To enhance the robustness and generalizability of our electric guitar transcription model, we employ content augmentation by leveraging an expanded version of the GuitarSet dataset as an augmentation resource, utilizing its acoustic guitar recordings to enrich the training data. The GuitarSet dataset \cite{xi2018guitarset} comprises acoustic-only guitar recordings, which we expanded using the same method applied to EGDB for creating EGDB-PG: re-rendering the audio with the Positive Grid BiasFX2 plugin to generate the same presets, resulting in 771 hours of rendered audio. During this process, we selected audio recorded using the pickup setting as the source to minimize the influence of microphone placement and environmental impulse responses, ensuring tone consistency when transforming acoustic recordings into electric guitar-like tones. This approach increases the tone diversity available for training, simulating a broader range of guitar sounds.

The use of the expanded GuitarSet dataset for content augmentation introduces a new strategy by combining acoustic and electric guitar data. As previously introduced, GuitarSet comprises acoustic guitar audio, while EGDB focuses on electric guitar audio. Previous studies have not combined these two types of data for training transcription models, often treating acoustic and electric guitar audio as separate domains due to their distinct acoustic properties. In our work, however, we utilize the expanded GuitarSet dataset as a content augmentation method, integrating its amplifier rendered augmented acoustic recordings with the electric guitar data in EGDB-PG to train our model. This strategy enriches the training data with diverse playing styles and tone variations, supporting generalization to electric guitar contexts.

\subsection{Audio nomarlization}
In neural effects modeling \cite{chen2024towards, steinmetz2021pyloudnorm}, decibel (dB) normalization ensures consistent peak amplitudes across clean audio  \( x \) and amplifier-rendered audio  \( x_{r} \). We apply piecewise normalization on every amplifier rendered audio segment to enforce a peak amplitude of --12 dBFS using the following equations:
\[
x = \frac{x}{\max(|x|)},
\quad
x = x * 10^{\left(-12.0 / 20.0\right)}.
\]

\begin{table*}

\centering
\caption{Evaluation of various settings in our tone-informed hFT-Transformer model, including tone removal, tone embeddings, content augmentation, and audio normalization, for low-gain, crunch, and high-gain transcription tasks on the EGDB-PG test split, compared to the hFT-Transformer baseline model\cite{toyama2023automatic}, which was trained from scratch on EGDB-PG and content augmentation.}
\label{tab:egdb_test}
\resizebox{\textwidth}{!}{%
\begin{tabular}{c|ccc|cccccc|cccccc|cccccc}
\toprule
\#  & \textbf{Model} & \textbf{Content} & \textbf{Norm} & \multicolumn{6}{c|}{\textbf{Low-gain}} & \multicolumn{6}{c|}{\textbf{Crunch}} & \multicolumn{6}{c}{\textbf{High-gain}} \\ 
 & & \textbf{Aug.} & & \multicolumn{3}{c}{Onset} & \multicolumn{3}{c|}{Frame} & \multicolumn{3}{c}{Onset} & \multicolumn{3}{c|}{Frame} & \multicolumn{3}{c}{Onset} & \multicolumn{3}{c}{Frame} \\
 & & & & F1 & P & R & F1 & P & R & F1 & P & R & F1 & P & R & F1 & P & R & F1 & P & R \\
\midrule
1  &TIT& & & 58.4 & 73.4 & 48.5 & 46.6 & 47.1 & 53.8 & 60.3 & 64.1 & 57.0 & 48.4 & 42.1 & 56.8 & 54.6 & 58.8 & 50.9 & 44.8 & 38.5 & 53.5 \\
2  & TIT&  & \checkmark & 71.3 & 69.0 & 73.8 & 63.3 & 56.1 & 72.6 & 69.9 & 62.5 & 79.4 & 60.3 & 52.2 & 71.3 & 67.1 & 75.0 & 60.7 & 55.8 & 47.8 & 67.1 \\

3  & TIT & \checkmark & & \textbf{84.0} & 81.5 & 86.6 & 68.4 & 59.4 & 80.1 & \textbf{81.6} & \textbf{77.6} & 86.0 & \textbf{67.6} & \textbf{58.7} & \textbf{79.8} & 78.9 & \textbf{75.5} & 82.6 & 64.5 & \textbf{56.5} & 75.3 \\
4  & TIT & \checkmark & \checkmark & 83.2 & 80.8 & 85.7 & \textbf{68.8} & \textbf{60.3} & \textbf{80.2} & 80.9 & 76.7 & 85.6 & 66.4 & 57.3 & 78.9 & \textbf{79.1} & 75.3 & 83.3 & \textbf{64.7} & 56.4 & \textbf{75.8} \\


5  & hFT\cite{toyama2023automatic} & \checkmark & & 79.6 & \textbf{85.2} & 74.7 & 63.6 & 55.3 & 74.9 & 78.8 & 74.3 & 83.9 & 62.6 & 54.9 & 72.9 & 76.8 & 72.8 & 81.4 & 61.1 & 53.9 & 70.4 \\

\bottomrule
\end{tabular}%
}

\end{table*}

\section{Experiments}
\label{sec:expereiment}

We first conduct an ablation study to assess the impact of different components, including tone embedding usage, training data augmentation (with the expanded GuitarSet dataset), and audio normalization, on transcription performance. We then compare TIT, trained with the best settings from the ablation study, against baseline models such as the hFT-Transformer~\cite{toyama2023automatic} and MT3 variants~\cite{gardner2021mt3}. Finally, we evaluate TIT on an unseen amplifier tone dataset, which includes tones not present in the training presets to test its robustness across diverse tone-related conditions.

To compute the spectrogram input, we apply a Hann window with a window size of 2048 and a Fast Fourier Transform (FFT) size of 2048, using 256 mel frequency bins (F), a constant padding mode, and a hop size of 256. The magnitude of the mel spectrogram is then compressed using a logarithmic function. We train our model with a batch size of 10 and a learning rate of 0.00005, using the Adam optimizer~\cite{kingma2014adam}, and follow the learning rate scheduling outlined in~\cite{toyama2023automatic}. All models are trained for 10,000 epochs on the EGDB-PG dataset, with some configurations including content augmentation as described in Section~\ref{methodology}. Training with content augmentation requires approximately 6 days to converge on a single NVIDIA V100 GPU. We select the best model for evaluation based on validation performance, ensuring optimal generalization to unseen amplifier types. Model performance is quantified using note-based metrics (focusing on onsets) and frame-level metrics, following the methodology in~\cite{hawthorne2017onsets}, with scores calculated on a per-piece basis and averaged to provide a comprehensive evaluation of transcription accuracy across amplifier-rendered audio.

\subsection{Impact of Content Augmentation and Audio Normalization on Transcription Performance}

In this section, we evaluate the performance of our models when trained on electric guitar audio from the EGDB-PG dataset, both with and without content augmentation. We first assess the baseline performance using only the EGDB-PG dataset (i.e., without content augmentation). We then explore the impact of incorporating the amplifier rendered GuitarSet as an augmentation dataset. The results, presented in Table~\ref{tab:egdb_test}, demonstrate the effects of content augmentation, audio normalization, and tone embeddings on transcription accuracy across various amplifier types.


\begin{figure*}[ht]
    \centering
    \subfloat[low-gain]{%
        \includegraphics[width=0.32\linewidth]{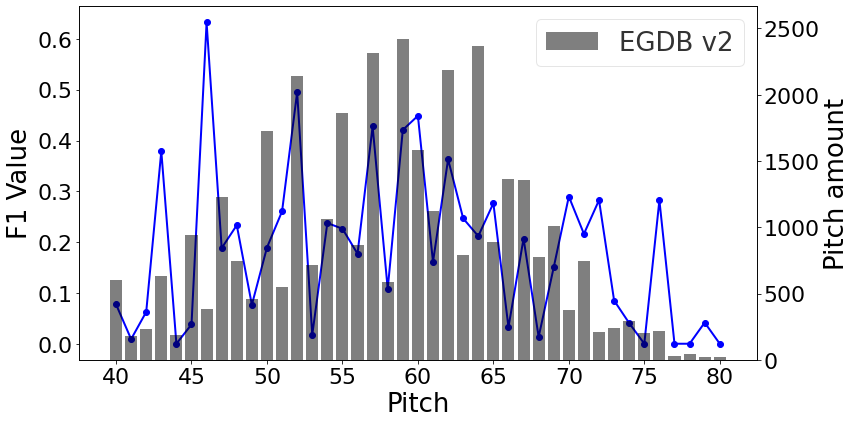}
        \label{fig:sub1}
    }
    \hfill
    \subfloat[crunch]{%
        \includegraphics[width=0.32\linewidth]{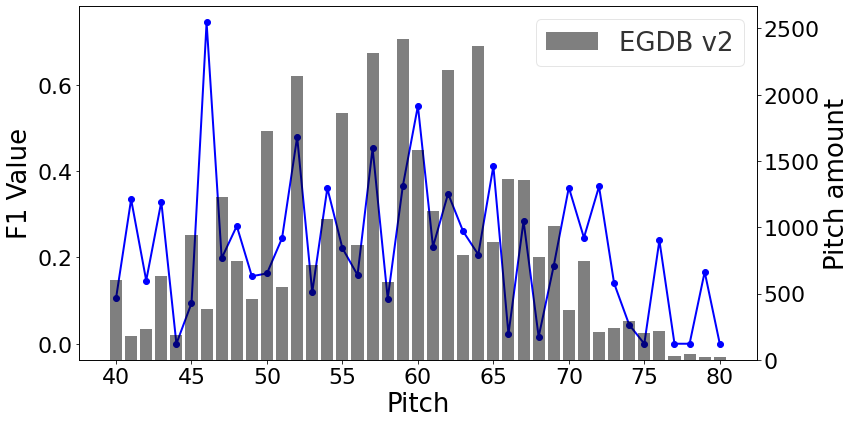}
        \label{fig:sub2}
    }
    \hfill
    \subfloat[high-gain]{%
        \includegraphics[width=0.32\linewidth]{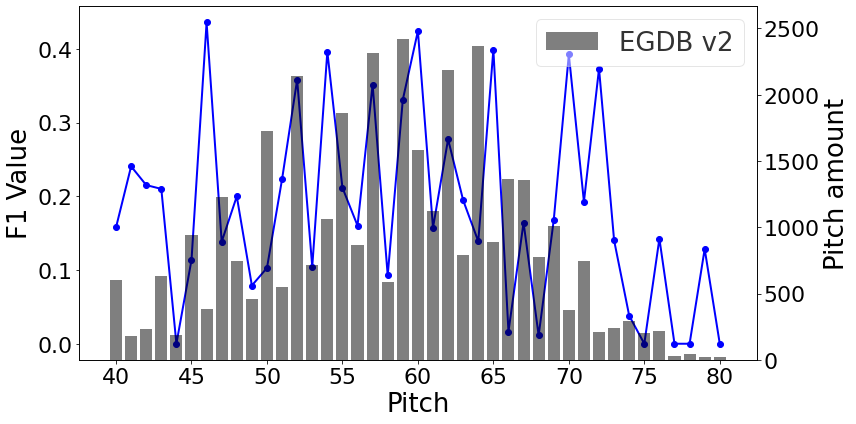}
        \label{fig:sub3}
    }
    \newline
    \subfloat[low-gain]{%
        \includegraphics[width=0.32\linewidth]{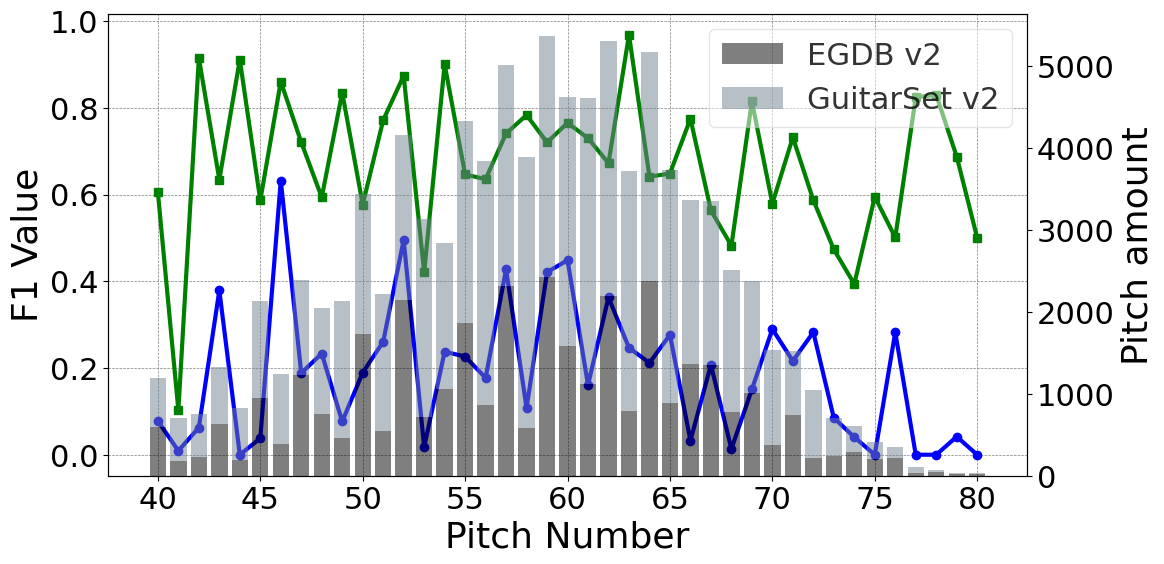}
        \label{fig:sub4}
    }
    \hfill
    \subfloat[crunch]{%
        \includegraphics[width=0.32\linewidth]{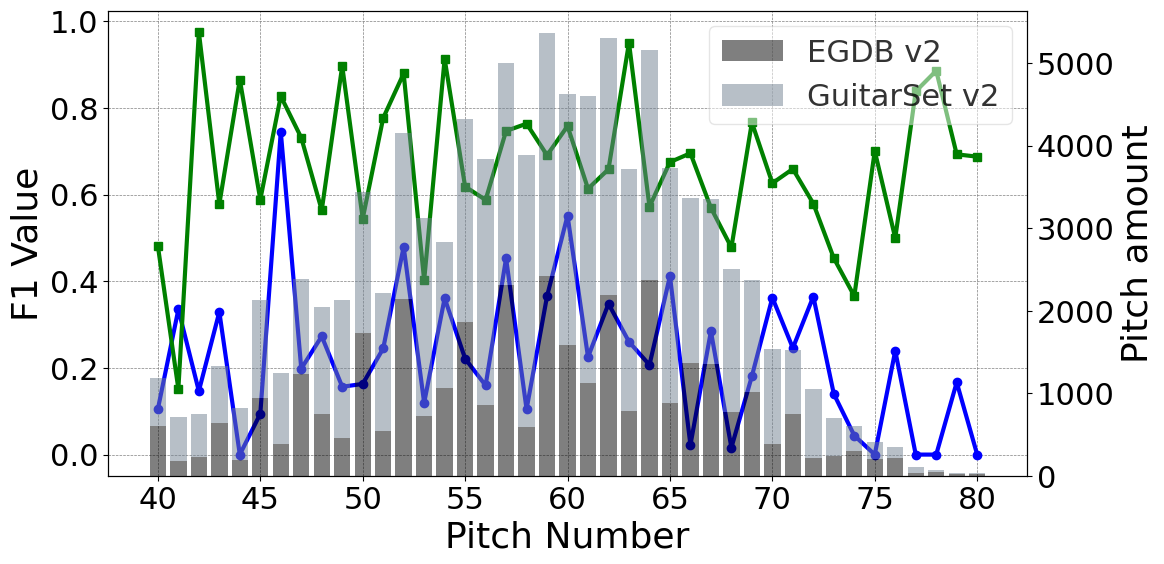}
        \label{fig:sub5}
    }
    \hfill
    \subfloat[high-gain]{%
        \includegraphics[width=0.32\linewidth]{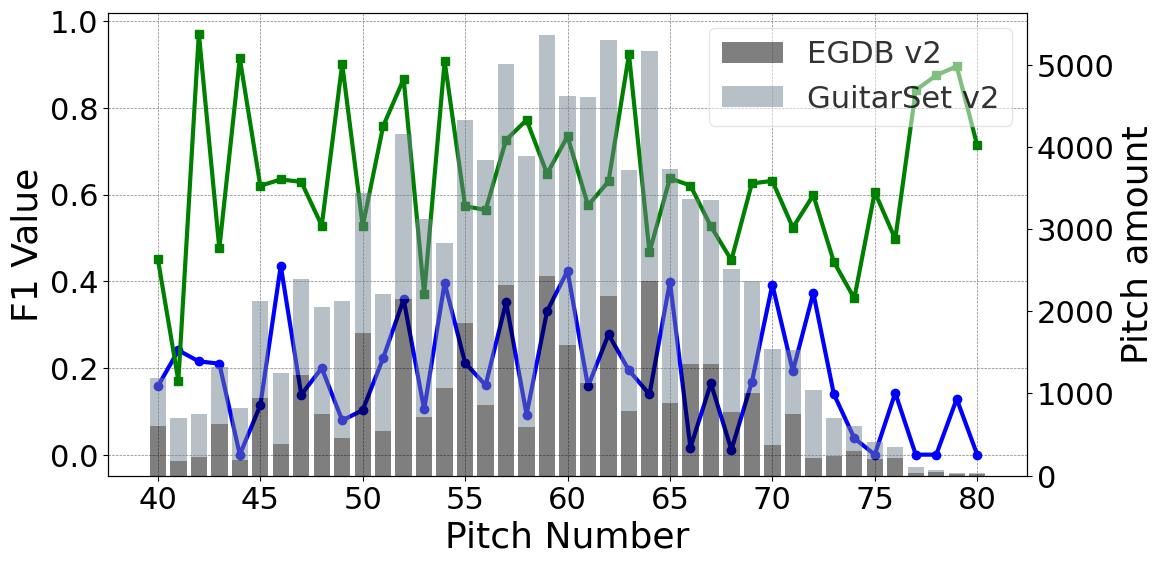}
        \label{fig:sub6}
    }

    \caption{Figure illustrating match note onset F1 values tested on the EGDB-PG test split. The blue line represents the TIT trained with EGDB-PG, while the green line represents the TIT trained with content augmentation, in addition to EGDB-PG. The histograms display the pitch distributions from EGDB-PG and amplifier rendered GuitarSet. The top row shows F1 values from the EGDB-PG only training approach across three amplifiers (low-gain, crunch, and high-gain). The bottom row compares models trained with EGDB-PG alone to those trained with both EGDB-PG and content augmentation. Results demonstrate that utilized content augmentation improves performance, especially for higher pitches, by leveraging the combined datasets.}
    \label{fig:pitch_score}
\end{figure*}

\subsubsection{Baseline Performance with Audio Normalization}

In rows 1 and 2 of Table~\ref{tab:egdb_test}, we show the impact of audio normalization on models trained solely on the EGDB-PG dataset without content augmentation. Applying dB normalization enhances both note-wise and frame-wise transcription performance across all amplifier types, particularly in low-resource training scenarios. For high-gain amplifiers, which present greater transcription challenges due to their complex harmonics, normalization yields significant improvements. The Onset F1 score rises by 12.9\%, 9.6\%, and 12.5\% for low-gain, crunch, and high-gain amplifier types, respectively. These results demonstrate the robustness of dB normalization in managing diverse amplifier tones when trained exclusively on the EGDB-PG dataset.

\subsubsection{Effect of Content Augmentation}

In rows 1 and 3 of Table~\ref{tab:egdb_test}, the comparison shows the effectiveness of content augmentation when using tone embeddings. Incorporating the amplifier-rendered GuitarSet into EGDB-PG creates a larger and more diverse training corpus, enhancing content variety and yielding improved note-wise and frame-wise scores across all amplifier types. Notably, in the challenging high-gain category, the note-wise F1 score increases from 54.6\% to 78.9\%. Further analysis of F1 scores for each pitch, presented in Figure~\ref{fig:pitch_score}, reveals substantial improvements for higher pitches (ranging from 70 to 80 MIDI notes, corresponding to high E to high G\# on a standard guitar). This improvement can be attributed to the expanded content diversity and the broader transformation of the rendering process from acoustic guitar audio to amplified settings, highlighting the effectiveness of content augmentation in improving the transcription of amplifier-rendered electric guitar audio.

\subsubsection{Impact of Audio Normalization under Content Augmentation}

In rows 3 and 4 of Table~\ref{tab:egdb_test}, we show the impact of audio normalization when content augmentation is applied. The benefits of audio normalization diminish compared to training on EGDB-PG alone, with F1 scores decreasing for low-gain and crunch amplifiers, while high-gain amplifier scores remain stable, improving by only 1.2\%. This reduction may stem from the time-axis Transformer in the second stage of TIT, which adapts to the enriched content diversity introduced by content augmentation. With sufficient training resources, such as the increased duration and variety provided by augmented content, audio normalization becomes less critical, as the model naturally generalizes across amplifier types.

\subsubsection{Performance without Tone Embeddings}

In rows 3 and 5 of Table~\ref{tab:egdb_test}, the comparison shows the effect of excluding tone embeddings while using content augmentation. We evaluate the performance of the hFT-Transformer~\cite{toyama2023automatic}, trained on the EGDB-PG dataset with the content augmentation. This setup achieves relatively high note-wise and frame-wise F1 scores across all amplifier types, underscoring the importance of incorporating datasets from different guitar types to enhance content diversity and transcription accuracy. However, adding tone embeddings further improves performance, increasing both note-wise and frame-wise F1 scores for all amplifier settings. This finding highlights the critical role of tone embeddings in enhancing transcription accuracy and consistency, particularly for capturing the complex tone-related variations in electric guitar audio.

\subsection{Comparison with other transcription model}

We evaluate our proposed model TIT alongside established baselines to assess their effectiveness in transcribing amplifier-rendered electric guitar audio. Our model, TIT, is trained from scratch on EGDB-PG with tone-informed embeddings and content augmentation, as our best configuration. In contrast, we test the following variants from the hFT-Transformer \cite{toyama2023automatic}:
\begin{itemize}
\item \textit{hFT-Maestro}: Pre-trained on Maestro\footnote{The released checkpoint and code for implementation are available at: \url{https://github.com/sony/hFT-Transformer}}, without additional training.
\item \textit{hFT-Maestro-EGDB-PG}: Pre-trained on Maestro, fine-tuned on EGDB-PG.
\end{itemize}

\begin{figure*}[ht]
    \centering
    \subfloat[log-mel spectrogram]{%
        \includegraphics[width=0.32\linewidth]{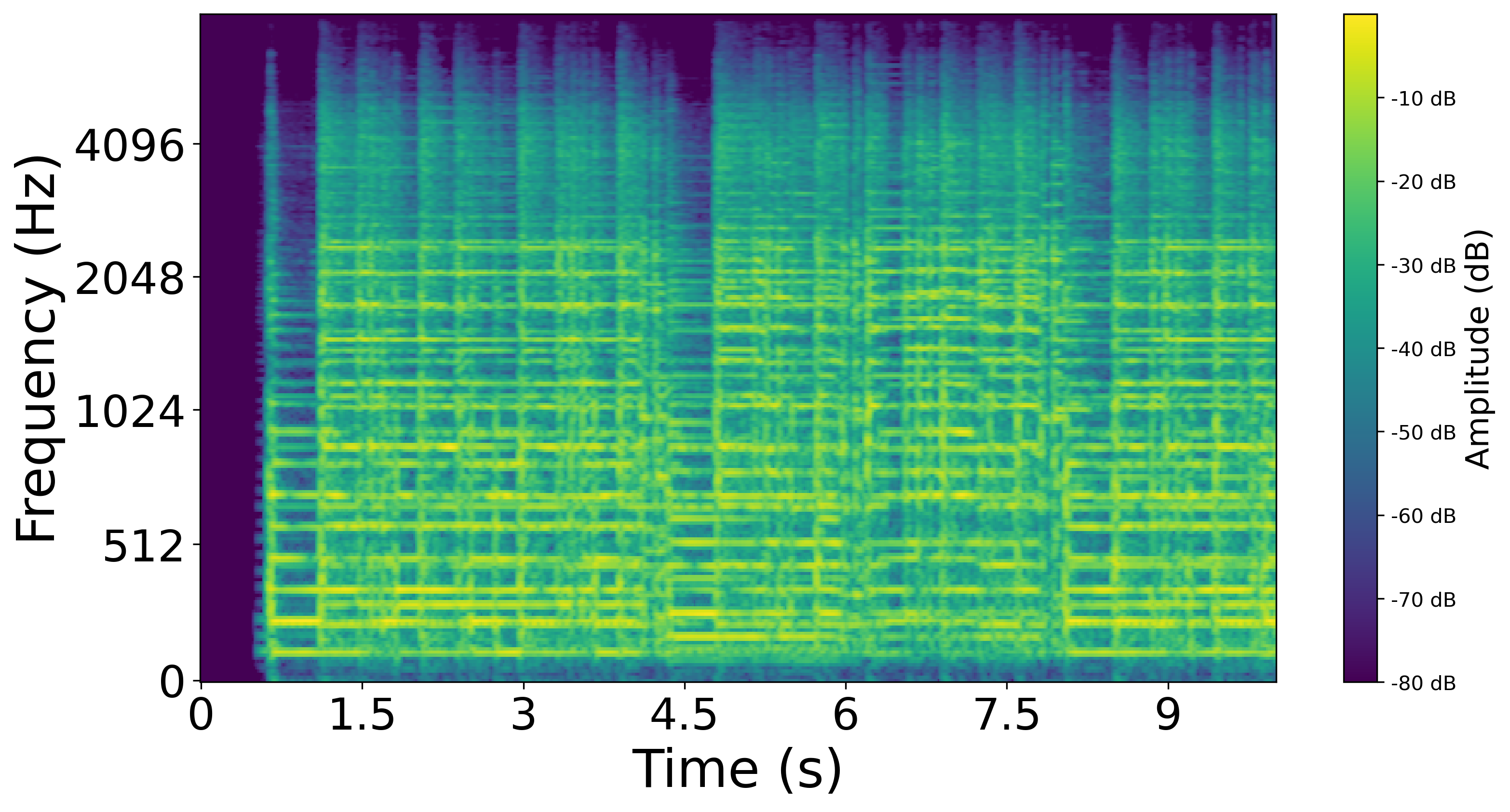}
        \label{fig:out_domain_sub1}
    }
    \hfill
    \subfloat[TIT]{%
        \includegraphics[width=0.32\linewidth]{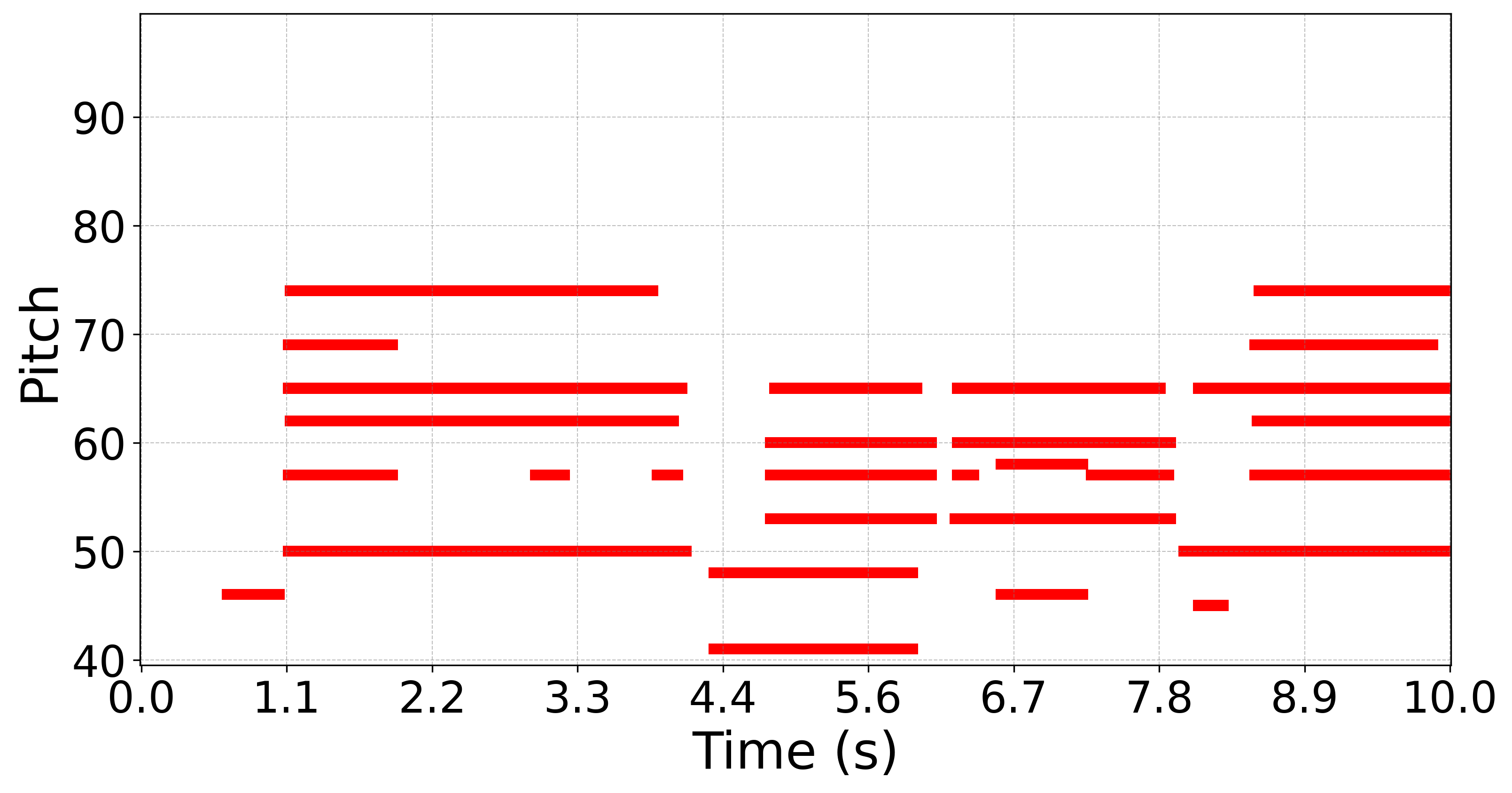}
        \label{fig:out_domain_sub2}
    }
    \hfill
    \subfloat[hFT-Maestro-EGDB-PG]{%
        \includegraphics[width=0.32\linewidth]{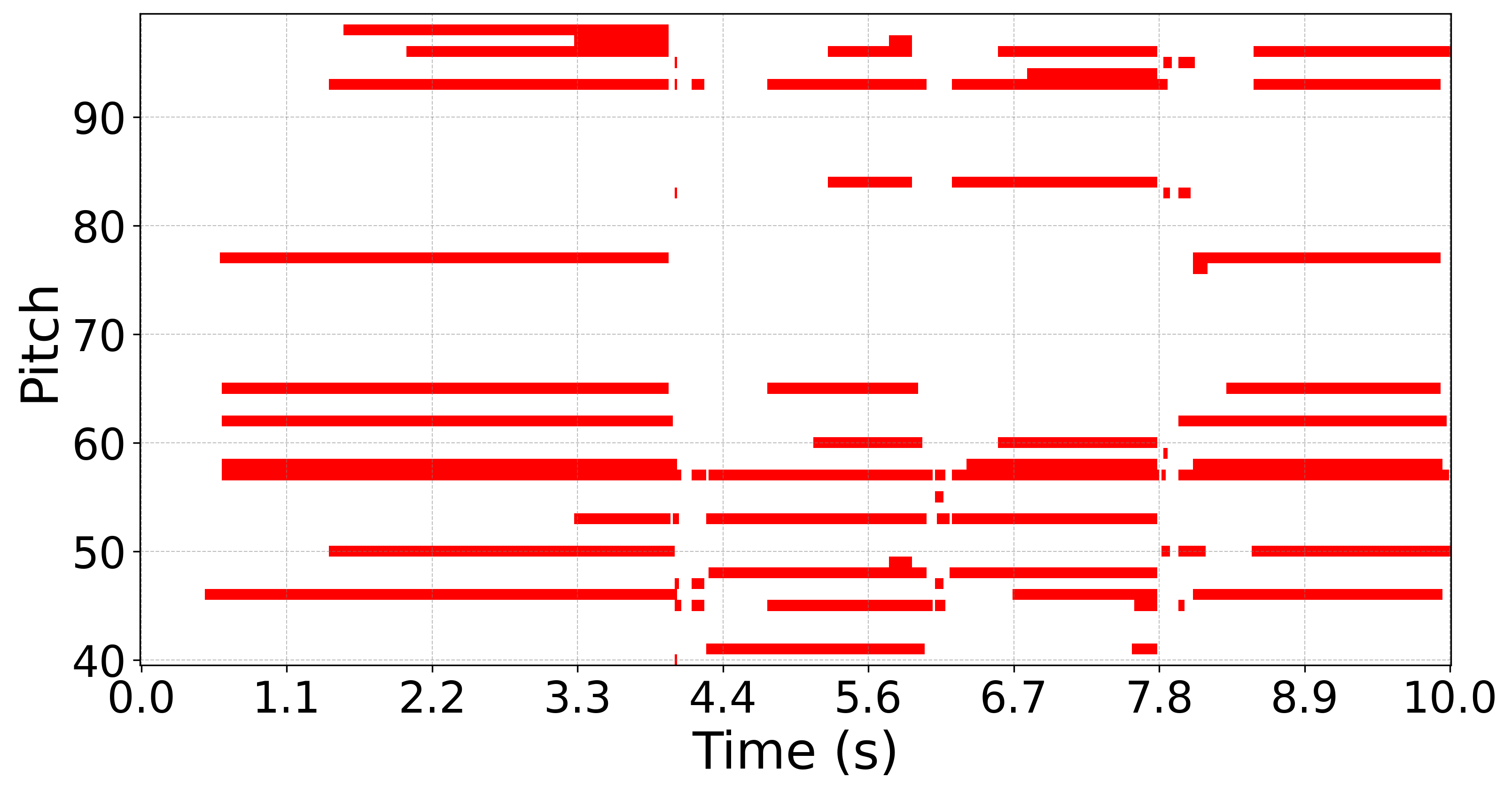}
        \label{fig:out_domain_sub3}
    }

    \caption{Illustration of out-of-domain 10-second audio clips. The leftmost (a) figure  displays the log-mel spectrogram representation of the audio. The middle (b) figure shows the piano roll predictions from the TIT model, while the rightmost (c) figure presents the piano roll predictions from the \textit{hFT-Maestro-EGDB-PG} model, finetuned on the EGDB-PG dataset. This comparison highlights the transcription performance of the two methods, with the \textit{hFT-M-finetuned} model generating notes that are often unplayable on a guitar, likely due to piano-specific trait from the Maestro dataset.} 
    \label{fig:out_domain_compared_vs_baseline}
\end{figure*}

\begin{table}


\centering
\caption{Performance evaluation of models on the EGDB-PG test split for low-gain, crunch, and high-gain transcription tasks. Results are compared with baselines (e.g., MT3) using onset and frame F1 scores.}
\label{tab:baseline}
\resizebox{\columnwidth}{!}{%
\begin{tabular}{l|cc|cc|cc}
\toprule
\textbf{Model} & \multicolumn{2}{c|}{\textbf{Low-gain}} & \multicolumn{2}{c|}{\textbf{Crunch}} & \multicolumn{2}{c}{\textbf{High-gain}} \\ 
& Onset F1 & Frame F1 & Onset F1 & Frame F1 & Onset F1 & Frame F1 \\
\midrule
\textit{TIT} & \textbf{84.0} & 68.4 & \textbf{81.6} & \textbf{67.6} & 78.9 & 64.5 \\
\textit{hFT-Maestro} & 74.4 & 54.5 & 47.5 & 42.2 & 42.6 & 39.1 \\
\textit{hFT-Maestro-EGDB-PG} & 59.6 & 52.2 & 52.2 & 43.4 & 50.1 & 41.5 \\
\textit{MT3-Guitar} & 44.2 & 7.9 & 12.8 & 9.9 & 46.8 & 7.8 \\
\textit{MT3-Guitar\&Piano} & 73.5 & 9.8 & 73.3 & 12.7 & 70.0 & 10.0 \\
\bottomrule
\end{tabular}%
}

\end{table}

To further assess the performance of multi-track transcription models, we incorporate the state-of-the-art MT3 model~\cite{gardner2021mt3} as a baseline for comparison. The MT3 model leverages a Transformer-based architecture to transcribe polyphonic audio across various instruments. We extract predictions from tracks labeled as guitar or piano to establish two baseline models, denoted as MT3-Guitar and MT3-Guitar\&Piano, respectively.\footnote{We inferred the transcription results from: \url{https://colab.research.google.com/github/magenta/mt3/blob/main/mt3/colab/music_transcription_with_transformers.ipynb}}

\begin{table*}[h!]

\centering
\caption{Evaluation of tone embeddings, content augmentation, and audio normalization in our tone embedding-informed hFT-Transformer model for low-gain, crunch, and high-gain transcription on out-of-domain test data, compared to the hFT-Transformer baseline model \cite{toyama2023automatic}, which was trained from scratch on EGDB-PG combined with content augmentation. Models with 256 tones and content augmentation (rows 1–2) outperform those with fewer tones (rows 5–8) and without content augmentation (rows 3–4), improving F1 scores. Audio normalization enhances performance in low-resource settings (rows 2, 4) but is less effective with content augmentation, highlighting its role in boosting accuracy for unseen amplifier tones.}

\label{tab:egdb_test_out_of_domain}
\resizebox{\textwidth}{!}{%
\begin{tabular}{c|cccc|cccccc|cccccc|cccccc}
\toprule
\# & \textbf{Model} &\textbf{\# of}  & \textbf{Content} & \textbf{Norm} & \multicolumn{6}{c|}{\textbf{Low-gain}} & \multicolumn{6}{c|}{\textbf{Crunch}} & \multicolumn{6}{c}{\textbf{High-gain}} \\ 
& & \textbf{Tones}& \textbf{Aug.}& & \multicolumn{3}{c}{Onset} & \multicolumn{3}{c|}{Frame} & \multicolumn{3}{c}{Onset} & \multicolumn{3}{c|}{Frame} & \multicolumn{3}{c}{Onset} & \multicolumn{3}{c}{Frame} \\
& & & & & F1 & P & R & F1 & P & R & F1 & P & R & F1 & P & R & F1 & P & R & F1 & P & R \\
\midrule
1 & TIT & 256 & \checkmark & & \textbf{84.8} & \textbf{87.2} & 82.6 & 69.3 & 63.4 & \textbf{76.4} & \textbf{85.3} & \textbf{83.8} & 86.8 & 70.0 & 62.1 & \textbf{80.2} & \textbf{79.0} & \textbf{76.8} & 81.5 & 65.7 & 58.5 & \textbf{75.0} \\
2 & TIT & 256  & \checkmark & \checkmark & 82.8 & 82.8 & \textbf{82.9} & \textbf{72.9} & \textbf{76.4} & 69.7 & 81.4 & 76.4 & \textbf{87.0} & \textbf{73.2} & \textbf{80.0} & 67.5 & 78.7 & 74.5 & \textbf{83.3} & \textbf{69.1} & \textbf{75.0} & 64.1 \\
\midrule
3 & TIT & 256  & & & 43.6 & 79.7 & 30.0 & 34.7 & 32.3 & 37.5 & 59.0 & 71.6 & 50.2 & 46.1 & 40.6 & 53.2 & 51.7 & 70.9 & 40.7 & 35.6 & 31.6 & 40.8 \\
4 & TIT & 256  & & \checkmark & 71.7 & 74.4 & 69.2 & 62.9 & 58.8 & 67.5 & 69.3 & 61.1 & 80.0 & 61.9 & 53.3 & 73.7 & 66.3 & 63.8 & 69.1 & 56.5 & 49.6 & 65.6 \\

5 & TIT & 1  & & & 23.4 & 40.8 & 16.4 & 11.9 & 11.1 & 12.9 & 20.7 & 66.1 & 12.3 & 9.2 & 7.0 & 13.6 & 2.5 & 27.2 & 1.3 & 7.3 & 6.3 & 8.8 \\
6 & TIT & 4  & & & 17.0 & 49.3 & 10.3 & 7.9 & 12.6 & 5.8 & 9.1 & 62.9 & 4.9 & 8.4 & 14.5 & 5.9 & 5.3 & 21.1 & 3.0 & 3.3 & 6.6 & 2.2 \\
7 & TIT & 1  & \checkmark & & 58.0 & 63.7 & 53.3 & 42.9 & 44.8 & 41.1 & 50.1 & 48.8 & 51.4 & 34.3 & 29.8 & 40.4 & 22.8 & 32.0 & 17.7 & 20.4 & 24.8 & 17.3 \\
8 & TIT & 4 & \checkmark & & 65.0 & 62.8 & 67.4 & 53.7 & 44.9 & 66.7 & 68.6 & 65.1 & 72.6 & 54.6 & 46.7 & 65.8 & 57.4 & 62.7 & 52.9 & 47.4 & 44.2 & 51.1 \\
\midrule
9 & hFT\cite{toyama2023automatic} & 256  & \checkmark & & 80.7 & 79.0 & 82.6 & 66.1 & 59.4 & 74.6 & 83.2 & 80.4 & 86.2 & 67.0 & 58.3 & 78.8 & 78.7 & 75.9 & 81.7 & 62.2 & 54.6 & 72.3 \\
\bottomrule
\end{tabular}%
}
\end{table*}

The performance of each model is presented in Table \ref{tab:baseline}. Among the hFT configurations, our TIT model delivers superior results, particularly in capturing tone diversity. It outperforms other hFT-Transformer variants across all metrics and amplifier types.

The \textit{hFT-Maestro model} excels in onset detection with low-gain amplifier settings, yet it struggles with crunch and high-gain amplifier types. This discrepancy highlights its limitations in transcribing electric guitar audio with high gain tones, where performance noticeably declines.

For the \textit{hFT-Maestro-EGDB-PG} model, we anticipated that fine-tuning the \textit{hFT-Maestro} model would improve F1 scores across all amplifier types. Surprisingly, transcription results reveal inconsistencies: while the model improve F1 score on crunch and high-gain settings, its F1 score degrades on low-gain audio. Closer analysis shows the model generates pitch notes uncharacteristic of guitar playing, likely due to piano-specific traits, such as left-hand comping and right-hand melody patterns, retained from its pre-training on the Maestro dataset. 
This leads to an overproduction of notes, reducing F1 score in both note-wise and frame-wise evaluations. We illustrate these transcription results in Figure \ref{fig:out_domain_compared_vs_baseline}.

For the MT3 baselines, using only the guitar track (\textit{MT3-Guitar}) yields lower note-wise precision and recall scores. Further investigation reveals that many note predictions are incorrectly assigned to the piano track (\textit{MT3-Guitar\&Piano}) instead. Despite this, our TIT model outperforms both MT3 variants in note-wise scores across all amplifier types. Frame-wise analysis further indicates that MT3 struggles to capture key features of amplified guitar audio, particularly in the later stages of the ADSR envelope, where performance weakens.

\subsection{Evaluating Generalizability to Tone and Content Variations}
In this section, we investigate the limitations of our models and each training mechanism. As noted in Section \ref{introduction}, the variety of presets and tones is nearly unlimited, making it essential to determine whether models trained on a dataset covering 256 tones can effectively transcribe unseen amplifier tones.

To assess the generalizability of our models to tone and content variations, we rendered audio from the EGDB-PG test set using six additional amplifiers from an external commercial plugin by the professional guitar amplifier company Neural DSP, creating audio with unseen amplifier tones. Following the categorical framework of EGDB-PG, these amplifier settings are classified into the same three types—low-gain, crunch, and high-gain—ensuring consistency with EGDB-PG.

To evaluate the impact of tone variations on the performance of transcribing electric guitar audio with unseen amplifier tones, we introduced two additional training configurations: ``1 tone`` and ``4 tones``. For the 1-tone setting, we used only the clean (DI) tone from the training data. For the 4-tone setting, we included the clean tone plus one representative tone from each amplifier type (totaling 1+3 tones). By varying the number of tones during training, we assess the effectiveness and performance of transcribing electric guitar audio. These modified models are detailed in rows 5–8 of Table \ref{tab:egdb_test_out_of_domain}.

We first examine the two best-performing models from Table \ref{tab:egdb_test} (rows 1 and 2), which demonstrate consistent performance for content augmentation, both with and without audio normalization. This stability reflects the models’ ability to adapt to diverse musical content on out-of-domain data. We also test models trained without content augmentation (rows 3 and 4 in Table \ref{tab:egdb_test_out_of_domain}), which exhibit lower performance compared to those using content augmentation, as shown in the same table. Additionally, we evaluate the approach without tone embeddings (row 4 in Table \ref{tab:egdb_test}), with results presented in row 9 of Table \ref{tab:egdb_test_out_of_domain}.

\subsubsection{Impact of Tone Augmentation (\# of Tones)} For tone variations, the number of tone settings used during training, we compare three configurations: training on DI tone audio alone (1 tone), training on DI tone audio augmented with one tone per amplifier type (4 tones), and training with 256 tones. 

The 256-tone models, enhanced by content augmentation (rows 1 and 2 in Table \ref{tab:egdb_test_out_of_domain}), outperform the 1-tone and 4-tone models (rows 5–8 in Table \ref{tab:egdb_test_out_of_domain}), also trained with content augmentation. Specifically, the 1-tone model (with augmentation) achieves a note-wise F1 score of 22.8\%, while the 4-tone model (with augmentation) achieves 57.4\%. Surprisingly, we initially hypothesized that the 4-tone model would outperform the clean-only model due to its exposure to greater tone diversity, even without content augmentation. However, the results (row 6 in Table \ref{tab:egdb_test_out_of_domain}) show slightly lower performance, with note-wise F1 scores decreasing for low-gain and crunch settings in the 4-tone model compared to the clean-only model. Analysis of the loss curve during training reveals that the 4-tone model converges rapidly, potentially leading to overfitting and reduced generalization on out-of-domain data.

\subsubsection{Impact of Content Augmentation} Pairwise comparisons in Table \ref{tab:egdb_test_out_of_domain} (row 1 vs row 3, row 2 vs row 4, row 5 vs row 7, and row 6 vs row 8) demonstrate that content augmentation improves all metrics (note-wise and frame-wise F1 scores) across all amplifier types in each setting. This validates the efficacy of content augmentation for enhancing model performance on unseen amplifier tones. Moreover, hft-Transformer (row 9 in Table \ref{tab:egdb_test_out_of_domain}) also shows desirable performance across all metrics, further confirming that employing content augmentation benefits the transcription of electric guitar audio with unseen amplifier tones, regardless of tone embedding use.

In summary, this study on out-of-domain transcription of audio with unseen amplifier tones reveals two key insights: 1) increasing the number of tones (tone augmentation) improves performance, but non-sufficient tone diversity (e.g., 4 tones) can lead to overfitting and reduced generalization compared to the 256-tone model, particularly without content augmentation; 2) content augmentation remains highly effective, consistently enhancing performance across all settings and amplifier types on out-of-domain data, even when tone embeddings or audio normalization are less impactful within this scope.

\section{Conclusion}
\label{conclusion}
To address the challenges of amplifier-rendered electric guitar transcription, we introduce EGDB-PG, a novel dataset capturing a wide range of tone variations, and propose the Tone-informed Transformer (TIT), a model enhanced with tone embeddings to improve adaptability to tone-related nuances. Our results establish the first comprehensive benchmark on EGDB-PG, highlighting the critical role of tone diversity in transcription tasks. Training techniques such as content augmentation, which increases content diversity and duration, along with tone embeddings and decibel normalization, originally developed for neural amplifier modeling, significantly enhance TIT’s transcription accuracy across both in-distribution and out-of-distribution samples, particularly training without content augmentation. Tone variety proves to be a pivotal factor, opening avenues for future research into broader effects and tone characteristics. Through experiments and ablation studies, this work advances the field by providing a robust benchmark for amplifier-rendered guitar transcription and demonstrating TIT’s sensitivity to tone variations, laying a strong foundation for future exploration of complex, effect-rendered transcription tasks on electric guitar.

\bibliographystyle{IEEEtran}

\bibliography{main}

\begin{thebibliography}{10}
\providecommand{\url}[1]{#1}
\csname url@samestyle\endcsname
\providecommand{\newblock}{\relax}
\providecommand{\bibinfo}[2]{#2}
\providecommand{\BIBentrySTDinterwordspacing}{\spaceskip=0pt\relax}
\providecommand{\BIBentryALTinterwordstretchfactor}{4}
\providecommand{\BIBentryALTinterwordspacing}{\spaceskip=\fontdimen2\font plus
\BIBentryALTinterwordstretchfactor\fontdimen3\font minus \fontdimen4\font\relax}
\providecommand{\BIBforeignlanguage}[2]{{%
\expandafter\ifx\csname l@#1\endcsname\relax
\typeout{** WARNING: IEEEtran.bst: No hyphenation pattern has been}%
\typeout{** loaded for the language `#1'. Using the pattern for}%
\typeout{** the default language instead.}%
\else
\language=\csname l@#1\endcsname
\fi
#2}}
\providecommand{\BIBdecl}{\relax}
\BIBdecl

\bibitem{emiya2009multipitch}
V.~Emiya, R.~Badeau, and B.~David, ``Multipitch estimation of piano sounds using a new probabilistic spectral smoothness principle,'' \emph{IEEE Transactions on Audio, Speech, and Language Processing (TASLP)}, 2009.

\bibitem{hawthorne2021sequence}
C.~Hawthorne, I.~Simon, R.~Swavely, E.~Manilow, and J.~Engel, ``Sequence-to-sequence piano transcription with transformers,'' \emph{International Society for Music Information Retrieval (ISMIR)}, 2021.

\bibitem{yan2021skipping}
Y.~Yan, F.~Cwitkowitz, and Z.~Duan, ``Skipping the frame-level: Event-based piano transcription with neural semi-crfs,'' \emph{Advances in Neural Information Processing Systems (NeurIPS)}, 2021.

\bibitem{ou2022exploring}
L.~Ou, Z.~Guo, E.~Benetos, J.~Han, and Y.~Wang, ``Exploring transformer’s potential on automatic piano transcription,'' in \emph{IEEE International Conference on Acoustics, Speech and Signal Processing (ICASSP)}, 2022.

\bibitem{hawthorne2017onsets}
C.~Hawthorne, E.~Elsen, J.~Song, A.~Roberts, I.~Simon, C.~Raffel, J.~Engel, S.~Oore, and D.~Eck, ``Onsets and frames: Dual-objective piano transcription,'' \emph{International Society for Music Information Retrieval (ISMIR)}, 2017.

\bibitem{kong2021high}
Q.~Kong, B.~Li, X.~Song, Y.~Wan, and Y.~Wang, ``High-resolution piano transcription with pedals by regressing onset and offset times,'' \emph{IEEE/ACM Transactions on Audio, Speech, and Language Processing (TASLP)}, 2021.

\bibitem{hawthorne2018enabling}
C.~Hawthorne, A.~Stasyuk, A.~Roberts, I.~Simon, C.-Z.~A. Huang, S.~Dieleman, E.~Elsen, J.~Engel, and D.~Eck, ``Enabling factorized piano music modeling and generation with the maestro dataset,'' \emph{International Conference on Learning Representations (ICLR)}, 2018.

\bibitem{toyama2023automatic}
K.~Toyama, T.~Akama, Y.~Ikemiya, Y.~Takida, W.-H. Liao, and Y.~Mitsufuji, ``Automatic piano transcription with hierarchical frequency-time transformer,'' \emph{International Society for Music Information Retrieval (ISMIR)}, 2023.

\bibitem{vaswani2017attention}
A.~Vaswani, ``Attention is all you need,'' \emph{Advances in Neural Information Processing Systems (NeurIPS)}, 2017.

\bibitem{wang2021soloist}
B.~Wang, M.~Y. Yang, and T.~Grossman, ``Soloist: Generating mixed-initiative tutorials from existing guitar instructional videos through audio processing,'' in \emph{Proceedings of the 2021 CHI Conference on Human Factors in Computing Systems}, 2021.

\bibitem{chen2022towards}
Y.-H. Chen, W.-Y. Hsiao, T.-K. Hsieh, J.-S.~R. Jang, and Y.-H. Yang, ``Towards automatic transcription of polyphonic electric guitar music: A new dataset and a multi-loss transformer model,'' in \emph{IEEE International Conference on Acoustics, Speech and Signal Processing (ICASSP)}, 2022.

\bibitem{lee2019audio}
J.~H. Lee, H.-S. Choi, and K.~Lee, ``Audio query-based music source separation,'' \emph{International Society for Music Information Retrieval (ISMIR)}, 2019.

\bibitem{liu2024separate}
X.~Liu, Q.~Kong, Y.~Zhao, H.~Liu, Y.~Yuan, Y.~Liu, R.~Xia, Y.~Wang, M.~D. Plumbley, and W.~Wang, ``Separate anything you describe,'' \emph{IEEE/ACM Transactions on Audio, Speech, and Language Processing (TASLP)}, 2024.

\bibitem{chen2022zero}
K.~Chen, X.~Du, B.~Zhu, Z.~Ma, T.~Berg-Kirkpatrick, and S.~Dubnov, ``Zero-shot audio source separation through query-based learning from weakly-labeled data,'' in \emph{Proceedings of the AAAI Conference on Artificial Intelligence (AAAI)}, 2022.

\bibitem{chen2024towards}
Y.-H. Chen, Y.-T. Yeh, Y.-C. Cheng, J.-T. Wu, Y.-H. Ho, J.-S.~R. Jang, and Y.-H. Yang, ``Towards zero-shot amplifier modeling: One-to-many amplifier modeling via tone embedding control,'' \emph{International Society for Music Information Retrieval (ISMIR)}, 2024.

\bibitem{xi2018guitarset}
Q.~Xi, R.~M. Bittner, J.~Pauwels, X.~Ye, and J.~P. Bello, ``Guitarset: A dataset for guitar transcription.'' in \emph{International Society for Music Information Retrieval (ISMIR)}, 2018.

\bibitem{manilow2019cutting}
E.~Manilow, G.~Wichern, P.~Seetharaman, and J.~Le~Roux, ``Cutting music source separation some slakh: A dataset to study the impact of training data quality and quantity,'' in \emph{IEEE Workshop on Applications of Signal Processing to Audio and Acoustics (WASPAA)}, 2019.

\bibitem{zang2024synthtab}
Y.~Zang, Y.~Zhong, F.~Cwitkowitz, and Z.~Duan, ``Synthtab: Leveraging synthesized data for guitar tablature transcription,'' in \emph{IEEE International Conference on Acoustics, Speech and Signal Processing (ICASSP)}, 2024.

\bibitem{sarmento2021dadagp}
P.~Sarmento, A.~Kumar, C.~Carr, Z.~Zukowski, M.~Barthet, and Y.-H. Yang, ``Dadagp: A dataset of tokenized guitarpro songs for sequence models,'' \emph{International Society for Music Information Retrieval (ISMIR)}, 2021.

\bibitem{wiggins2019guitar}
A.~Wiggins and Y.~E. Kim, ``Guitar tablature estimation with a convolutional neural network.'' in \emph{International Society for Music Information Retrieval (ISMIR)}, 2019.

\bibitem{su2019tent}
T.-W. Su, Y.-P. Chen, L.~Su, and Y.-H. Yang, ``Tent: Technique-embedded note tracking for real-world guitar solo recordings,'' \emph{Transactions of the International Society for Music Information Retrieval (TISMIR)}, 2019.

\bibitem{huang2023note}
T.-S. Huang, P.-C. Yu, and L.~Su, ``Note and playing technique transcription of electric guitar solos in real-world music performance,'' in \emph{IEEE International Conference on Acoustics, Speech and Signal Processing (ICASSP)}, 2023.

\bibitem{riley2024high}
X.~Riley, D.~Edwards, and S.~Dixon, ``High resolution guitar transcription via domain adaptation,'' in \emph{IEEE International Conference on Acoustics, Speech and Signal Processing (ICASSP)}, 2024.

\bibitem{abreu2024leveraging}
H.~P.~W. Abreu, R.~Corey, and I.~Roman, ``Leveraging electric guitar tones and effects to improve robustness in guitar tablature transcription modeling,'' \emph{International Conference on Digital Audio Effects (DAFx)}, 2024.

\bibitem{pedroza2022egfxset}
H.~Pedroza, G.~Meza, and I.~R. Roman, ``Egfxset: Electric guitar tones processed through real effects of distortion, modulation, delay and reverb,'' \emph{ISMIR Late Breaking Demo}, 2022.

\bibitem{stein2010automatic}
M.~Stein, J.~Abe{\ss}er, C.~Dittmar, and G.~Schuller, ``Automatic detection of audio effects in guitar and bass recordings,'' in \emph{Audio Engineering Society Convention 128 (AES)}, 2010.

\bibitem{chen2020simple}
T.~Chen, S.~Kornblith, M.~Norouzi, and G.~Hinton, ``A simple framework for contrastive learning of visual representations,'' in \emph{International conference on machine learning (ICML)}, 2020.

\bibitem{caron2020unsupervised}
M.~Caron, I.~Misra, J.~Mairal, P.~Goyal, P.~Bojanowski, and A.~Joulin, ``Unsupervised learning of visual features by contrasting cluster assignments,'' \emph{Advances in neural information processing systems (Neurips)}, 2020.

\bibitem{assran2022maskedsiamesenetworkslabelefficient}
\BIBentryALTinterwordspacing
M.~Assran, M.~Caron, I.~Misra, P.~Bojanowski, F.~Bordes, P.~Vincent, A.~Joulin, M.~Rabbat, and N.~Ballas, ``Masked siamese networks for label-efficient learning,'' 2022. [Online]. Available: \url{https://arxiv.org/abs/2204.07141}
\BIBentrySTDinterwordspacing

\bibitem{pedalboard2021}
\BIBentryALTinterwordspacing
P.~Sobot, ``pedalboard,'' 2021. [Online]. Available: \url{https://github.com/spotify/pedalboard}
\BIBentrySTDinterwordspacing

\bibitem{steinmetz2021pyloudnorm}
C.~J. Steinmetz and J.~D. Reiss, ``pyloudnorm: {A} simple yet flexible loudness meter in python,'' in \emph{150th AES Convention}, 2021.

\bibitem{gardner2021mt3}
J.~Gardner, I.~Simon, E.~Manilow, C.~Hawthorne, and J.~Engel, ``Mt3: Multi-task multitrack music transcription,'' \emph{International Conference on Learning Representations (ICLR)}, 2022.

\bibitem{kingma2014adam}
D.~P. Kingma, ``Adam: A method for stochastic optimization,'' \emph{International Conference on Learning Representations (ICLR)}, 2015.

\end{thebibliography}

\vfill

\end{document}